\documentclass[apj, twocolappendix, numberedappendix, appendixfloats]{emulateapj}
\usepackage[latin2]{inputenc}
\usepackage{amsmath}
\usepackage{graphicx}
\usepackage[pdftex,backref,breaklinks,colorlinks,citecolor=blue]{hyperref}
\usepackage{url}
\usepackage[pass,letterpaper]{geometry}

\begin{document}
\title{Gaia reveals a metal-rich in-situ component of the local stellar halo}
\journalinfo{The Astrophysical Journal, submitted}
\author{Ana Bonaca\altaffilmark{1}$^\dagger$}\thanks{$^\dagger$ ITC Fellow}
\author{Charlie Conroy\altaffilmark{1}}
\author{Andrew Wetzel\altaffilmark{2,3,4}$^\ddagger$}\thanks{$^\ddagger$ Caltech-Carnegie Fellow}
\author{Philip F. Hopkins\altaffilmark{2}}
\author{Du\v san Kere\v s\altaffilmark{5}}

\altaffiltext{1}{Department of Astronomy, Harvard University, Cambridge, MA 02138; {ana.bonaca@cfa.harvard.edu}}
\altaffiltext{2}{TAPIR, California Institute of Technology, Pasadena, CA, USA}
\altaffiltext{3}{Carnegie Observatories, Pasadena, CA, USA}
\altaffiltext{4}{Department of Physics, University of California, Davis, CA, USA}
\altaffiltext{5}{Department of Physics, Center for Astrophysics and Space Sciences, University of California, San Diego, La Jolla, CA, USA}

\begin{abstract}
We use the first \emph{Gaia} data release, combined with RAVE and APOGEE spectroscopic surveys, to investigate the origin of halo stars within $\lesssim3\;$kpc from the Sun.
We identify halo stars kinematically, as moving with a relative speed of at least 220\;km/s with respect to the local standard of rest.
These stars are in general more metal-poor than the disk, but surprisingly, half of our halo sample is comprised of stars with $\rm[Fe/H]>-1$.
The orbital directions of these metal-rich halo stars are preferentially aligned with the disk rotation, in sharp contrast with the isotropic orbital distribution of the more metal-poor halo stars.
We find similar properties in the Latte cosmological zoom-in simulation of a Milky Way-like galaxy from the FIRE project.
In Latte, metal-rich halo stars formed primarily inside of the solar circle, while lower-metallicity halo stars preferentially formed at larger distances (extending beyond the virial radius).
This suggests that metal-rich halo stars in the Solar neighborhood in fact formed in-situ within the Galactic disk rather than having been accreted from satellite systems.
These stars, currently on halo-like orbits, therefore have likely undergone substantial radial migration/heating.
\end{abstract}
\maketitle

\section{Introduction}
Early works suggested that the Galactic stellar halo formed in a dissipative collapse of a single protogalactic gas cloud \citep{els}, but in today's prevalent cosmology, galaxies grow hierarchically by accreting smaller structure \citep{white1978, diemand2008, springel2008, klypin2011}.
The archeological record of these galactic building blocks is especially well retained in stellar halos, in part because of their long relaxation times.
Upon accretion, smaller galaxies completely disrupt within several orbital periods \citep{helmi1999b}, and become a part of the smooth halo.
Still, global halo properties contain information on the original systems.
For example, the total mass of a halo is related to the number of massive accretion events \citep{bj2005, cooper2010}, the outer slope of the outer halo density profile increases with the current accretion rate \citep{diemer2014}, while the presence of a break in the density profile is indicative of a quiet merger history at late times \citep{deason2013}.
Therefore, when studying the stellar halo of a galaxy, we are in fact analyzing a whole population of accreted systems.

Modern works suggest that there are two distinct processes that contribute to the buildup of a stellar halo: most of the halo stars have been \emph{accreted} from smaller galaxies, but a fraction has been formed \emph{in situ} inside of the main galaxy \citep{zolotov2009, font2011, cooper2015}.
The in-situ component can contain stars formed in the initial gas collapse \citep{samland2003} and/or stars formed in the disk, which have subsequently been kicked out and placed on halo orbits \citep{purcell2010}.
Throughout this work we use the term in-situ halo to address both of these origin scenarios.
Numerical simulations show that the total number of halo stars formed in situ depends on the details of the formation history, but in general, their contribution decreases with distance from the galactic center \citep[e.g.,][]{zolotov2009, cook2016}.
For Milky Way-like galaxies, $10-30$\% of halo stars in the Solar neighborhood are expected to have formed in situ \citep{zolotov2009}. 
Thorough understanding of stellar halos relies on disentangling its accreted and in-situ components.

Recent accretion events are still coherent in configuration space, and have been observed throughout the Local Universe. 
The first evidence of accretion came from studies of globular clusters \citep{sz}, but in recent years deep photometric surveys on large scales have revealed a plethora of systems undergoing tidal disruption in the Milky Way halo, with progenitors ranging from dwarf galaxies \citep[e.g.,][]{ibata1994, belokurov2007a, belokurov2007b, juric2008, bonaca2012a} to disrupted globular clusters \citep[e.g.,][]{rockosi2002, grillmair2006, grillmair2009, bonaca2012b, bernard2016}.
A complete list of tidal structures identified in the Milky Way has been recently compiled by \citet{grillmair2016}.
Extragalactic surveys of low surface brightness features indicate that ongoing accretion is common in present-day galaxies of all masses \citep{ibata2001, ferguson2002, martinez-delgado2010, martinez-delgado2012, romanowsky2012, crnojevic2016}.

Evidence of past accretion events remains even after they no longer stand out as overdensities in configuration space.
For example, \citet{bell2008} measured that the spatial distribution of halo stars in the Milky Way is highly structured, at the degree expected by models where the entire halo was formed by accretion of dwarf galaxies.
Expanding this approach to clustering in the space of both positions and radial velocities, \citet{janesh2016} found that the amount of halo structure grows with distance from the Galactic center.
However, phase-space substructure associated with mergers has been discovered even in the Solar neighborhood \citep{helmi1999,smith2009,helmi2017}.

While there are multiple avenues for identifying accreted stars in a halo, isolating an in-situ component has been more challenging.
Early observations of globular clusters and individual stars indicated that the inner halo is more metal-rich, has a metallicity gradient, and is slightly prograde, while objects in the outer halo are more metal-poor and retrograde with respect to the Galactic disk \citep[e.g.,][]{sz}.
This has been interpreted as evidence that the inner halo is formed in situ, while the outer halo is accreted.
With the advent of wide-field sky surveys, these findings of a dual stellar halo have been confirmed using much larger samples of halo stars \citep{carollo2007, carollo2010, beers2012}.
Adding further evidence for a presence of an in-situ component, \citet{schlaufman2012} ruled out accretion as the main origin of stars in the inner halo due to their lack of spatial coherence with metallicity.
Most recently, \citet{deason2017} measured a small rotation signal among the old halo stars, which is consistent with the halo having a minor in-situ component.

Even though there is abundant evidence that both in-situ and accreted stars are present in the Milky Way halo, their contributions haven't yet been properly accounted for.
A straightforward way to distinguish between these two origin scenarios would be to directly compare halo stars in the Milky Way to a simulated halo, where the origin of every star is known.
This comparison is most easily performed by matching stars on their orbital properties, but precise observations of halo stars that would allow such a match have so far been limited to a distance of a few hundred pc from the Sun -- a volume poorly resolved in hydrodynamical simulations of large galaxies such as the Milky Way.
Fortunately, major improvements recently occurred on both the observational and theoretical fronts.
The \emph{Gaia} mission \citep{perryman2001} has increased the available observed volume by an order of magnitude.
Furthermore, \emph{Gaia} measurements are much more precise than previously available data, whose role in establishing the presence of a dual halo drew some criticism \citep{schonrich2011, schonrich2014}.
On the theory side, \citet{wetzel2016} recently presented the Latte high-resolution cosmological zoom-in simulation of a Milky Way-mass galaxy.
We leverage the joint powers of the new \emph{Gaia} data set and the Latte simulation to reveal the origin of the stellar halo in the Solar neighborhood.

This paper is organized as follows: we start by introducing our observational data that combine the first year of \emph{Gaia} astrometry with ground-based radial velocity measurements in \S\ref{sec:data}.
Once we've compiled a data set that has 6D information for stars in the Solar neighborhood, we split the sample into a disk and a halo component in \S\ref{sec:halostars} and analyze their chemical abundances and orbital properties.
We interpret these observations in terms of a simple toy model, as well as using the Latte cosmological hydrodynamic simulation of a Milky Way-like galaxy in \S\ref{sec:origin}, where we also propose a formation scenario for kinematically selected halo stars close to the Sun.
Section \S\ref{sec:discussion} discusses broader implications of our findings, which are summarized in \S\ref{sec:summary}.

\begin{figure*}
\begin{center}
\includegraphics[width=\textwidth]{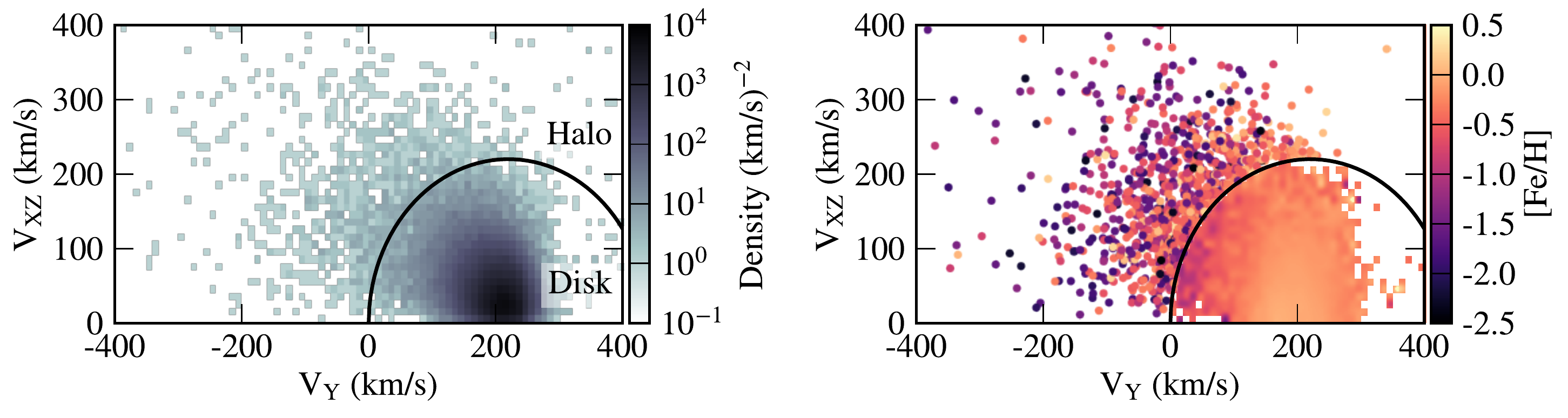}
\caption{(Left) Toomre diagram of stars in the Solar neighborhood, from a combined catalog of \emph{Gaia}--TGAS proper motions and parallaxes, and RAVE-on radial velocities, thus covering the full 6D phase space.
We kinematically divide the sample into a disk and a halo component.
The halo stars are defined as having $|V-V_{LSR}|>220$\,km/s, and the dividing line is shown in black.
(Right) Positions of TGAS--RAVE-on stars with a measured metallicity in the Toomre diagram.
The color-coding corresponds to the average metallicity of stars in densely populated regions of the diagram, and individual metallicities otherwise.
Interestingly, many halo stars are metal-rich, with $\rm[Fe/H]>-1$.}
\label{fig:toomre}
\end{center}
\end{figure*}

\section{Data}
\label{sec:data}
Studying orbital properties in a sample of stars requires the knowledge of their positions in 6D phase space.
Currently, \emph{Gaia} Data Release 1 (DR1) provides the largest and most precise 5D dataset for stars in the Solar neighborhood, which we describe in \S\ref{gaia}.
We complement this data with radial velocities from ground-based spectroscopic surveys whose targets overlap with \emph{Gaia} (\S\ref{rvsurveys}).
Finally, we describe our sample selection in \S\ref{sample}.

\subsection{Gaia}
\label{gaia}
\emph{Gaia} is a space-based mission that will map the Galaxy over the next several years \citep{perryman2001}.
The first data from the mission was released in September 2016, and contains not only positions of all \emph{Gaia} sources ($G<20$), but also positions, parallaxes and proper motions for $\sim$2~million of the brightest stars in the sky \citep{gaiadr1, gaiamission}.
Obtaining the 5D measurements after just a year of \emph{Gaia}'s operation was possible by referencing the positions measured with Hipparcos \citep{michalik2015}.
The faintest stars observed by Hipparcos \citep{hipparcos, vleeuwen2007} and released as a part of Tycho~II catalog have $V\sim12$ \citep{hog2000}, which limits the size of the 5D sample in \emph{Gaia} DR1 to $\approx2$ million stars.
The joint solution, known as Tycho--Gaia Astrometric Solution \citep[TGAS,][]{gaiaastrometry}, is comparable in attained proper motions and parallaxes to the Hipparcos precision (typical uncertainty in positions and parallaxes is 0.3\;mas and 1\;mas/yr in proper motions), but already on a sample that is more than an order of magnitude larger, making TGAS an unrivaled dataset for precision exploration of the Galactic phase space.

\subsection{Spectroscopic surveys}
\label{rvsurveys}
\emph{Gaia} is measuring radial velocities for $\sim150,000$ stars brighter than $G<16$ \citep{gaiamission}, but the first spectroscopic data will become available only in the second data release.
Thus, we completed the phase-space information of TGAS sources by using radial velocities from ground-based spectroscopic surveys.
We used two distinct spectroscopic datasets, from the RAVE and APOGEE projects, and provide an overview below.

The Radial Velocity Experiment \citep[RAVE,][]{steinmetz2006} is a spectroscopic survey of the southern sky, and its magnitude range $9<I<12$ is well matched to TGAS.
The latest data release, RAVE DR5 \citep{kunder2017}, contains $\sim450,000$ unique radial velocity measurements.
Since RAVE avoided targeting regions of low galactic latitude, the actual overlap with TGAS is $\sim250,000$ stars -- the largest of any spectroscopic survey.
The survey was performed at the UK Schmidt telescope with the 6dF multi-object spectrograph \citep{6df}, in the wavelength range $8410-8795\,\rm\AA$ at a medium resolution of $R\sim7,500$, so the typical velocity uncertainty is $\sim2$\,km/s.
Abundances of up to seven chemical elements are available for a subset of high signal-to-noise spectra.
\citet{casey2016} reanalyzed the RAVE DR5 spectra in a data-driven fashion with The Cannon \citep{ness2015}, providing de-noised measurements of stellar parameters and chemical abundances in the RAVE-on catalog.
In particular, typical uncertainty in RAVE-on abundances is 0.07\;dex, which is at least 0.1\;dex better than precision achieved using the standard spectroscopic pipeline.
Therefore, we opted to use RAVE-on chemical abundances, focusing on metallicities, [Fe/H], and $\alpha$-element abundances.

The Apache Point Observatory Galactic Evolution Experiment (APOGEE) is one of the programs in the Sloan Digital Sky Survey III \citep{majewski2015, sdss3}, which acquired $\sim500,000$ infrared spectra for $\sim150,000$ stars brighter than $H\sim12.2$ \citep{holtzman2015}.
To capitalize on the infrared wavelength coverage, APOGEE mainly targeted the disk plane, but several high latitude fields are included as well \citep{zasowski2013}.
Its higher resolution $R\sim22,500$, provides more precise abundances for a larger number of elements \citep[e.g.,][]{ness2015}.
APOGEE targets are preferentially fainter than stars targeted by RAVE, so its overlap with TGAS is limited to a few thousand stars.
APOGEE and RAVE have different footprints, targeting strategy, and the wavelength window observed, so despite the smaller sample size, we found APOGEE to be a useful dataset for validating conclusions reached by analyzing the larger RAVE sample.

\subsection{Sample selection}
\label{sample}
After matching Gaia--TGAS to the spectroscopic surveys, we increase the quality of the sample by excluding stars with large observational uncertainties.
However, the overlap between TGAS and spectroscopic surveys is limited, and the number density of halo stars in the Solar neighborhood is low.
To ensure that we have a sizeable halo sample, we chose to use very generous cuts on observational uncertainties and propagate them when interpreting our results, rather than restricting our sample size by more stringent cuts.
In particular, we included stars with radial velocity uncertainties smaller than 20\,km/s, and relative errors in proper motions and parallaxes smaller than 1.
In addition, we removed all stars with a negative parallax, to simplify the conversion to their distance.
These criteria select 159,352 stars for the TGAS--RAVE-on dataset, and 14,658 stars for the TGAS--APOGEE sample.

The spatial distribution of stars in our sample is entirely determined by the joint selection function of TGAS and the spectroscopic surveys, as we performed no additional spatial selection.
\emph{Gaia} is an all-sky survey, but since the data is still being acquiring, completeness of the TGAS catalog varies across the sky.
Ground-based spectroscopic surveys have geographically restricted target-list, in addition to the adopted targeting strategy.
This results in a spatially non-uniform sample, whose distribution in distances we provide in Appendix~\ref{sec:distances}.
On the other hand, \citet{wojno2016} have shown that the RAVE survey is both chemically and kinematically unbiased.
Thus, focusing on kinematic properties of the sample will result in robust conclusions.

\section{Properties of the local halo stars}
\label{sec:halostars}
In this section we analyze the properties of halo stars in a sample of bright stars, $V\lesssim12$, within 3\;kpc from the Sun, that have positions, proper motions and parallaxes in the TGAS catalog, and radial velocities from either RAVE or APOGEE.
This sample, though spatially incomplete due to survey selection functions, is kinematically unbiased.
We define a kinematically-selected halo in \S~\ref{sec:sample}, and present its chemical and orbital properties in \S~\ref{sec:chem} and \S~\ref{sec:l}, respectively.

\subsection{Defining a local sample of halo stars}
\label{sec:sample}
Access to the full 6D phase space information allows us to calculate Galactocentric velocities for all of the stars in the sample.
We summarize the kinematic properties of the sample with a Toomre diagram (Figure~\ref{fig:toomre}), where the Galactocentric $Y$ component on the velocity vector, $V_Y$ (which points in the direction of the disk rotation), is on the x axis, while the perpendicular Toomre component, $\sqrt{V_X^2+V_Z^2}$, is on the y axis. 
This space has been widely used to identify major components of the Galaxy: the thin and thick disks, and the halo \citep[e.g.,][]{venn2004}.
Disk stars dominate a large overdensity at $V_Y\approx220$\;km/s, which corresponds to the circular velocity of the Local Standard of Rest (LSR, $V_{LSR}$).
The density of stars (left panel of Figure~\ref{fig:toomre}) decreases smoothly for velocities progressively more different from $V_{LSR}$, extending all the way to retrograde orbits ($V_Y<0$).

We distinguish between the disk and the halo following \citet{ns2010}: halo stars are identified with a velocity cut $|V-V_{LSR}|>220$\;km/s, where $V_{LSR} = (0,220,0)$\;km/s in the Galactocentric Cartesian coordinates.
The dividing line between the components is marked with a black line in Figure~\ref{fig:toomre}, and both components are labeled in the left panel.
The halo definition employed here is more conservative than similar cuts adopted by previous studies; e.g., \citet{ns2010} defined halo as stars with velocities that satisfy $|V-V_{LSR}|>180$\;km/s.
For example, \citet{sb2009} have shown that the velocity distribution of a Galactic thick disk can be asymmetric, in which case the region $180<|V-V_{LSR}|<220$\;km/s could still contain many thick disk stars.
A higher velocity cut ensures that the contamination of our halo sample with thick disk stars is minimized.
In total, we identified 1,376 halo and 157,976 disk stars, with the halo constituting $\sim1\%$ of our sample.
This is in line with the expectations from number count studies in large-scale surveys \citep[e.g.,][]{juric2008}, although we do not expect an exact match, as TGAS is not volume complete.

\subsection{Chemical composition}
\label{sec:chem}
In this section we study the chemical composition of the Solar neighborhood stars observed by both Gaia--TGAS and RAVE.
The signal-to-noise ratio of 142,086 RAVE spectra was high enough to allow a measurement of metallicity [Fe/H].
Alpha-element abundances, [$\alpha$/Fe], were obtained for a subset of 56,259 stars.
Right panel of Figure~\ref{fig:toomre} shows the average metallicity in densely populated velocity bins of the Toomre diagram, while the points in the lower density regions are individually colored-coded by $\rm[Fe/H]$.
As expected, the halo is more metal poor than the disk \citep[e.g.,][]{gilmore1989, ivezic2008}.
Within the disk itself, there is a smooth decrease in metallicity further from the $V_{LSR}$, starting from [Fe/H]$\sim0$ in the thin disk region, $(V_Y, V_{XZ})=(220,0)$\;km/s, to $\rm[Fe/H]\sim-0.5$ in the thick disk region, $(V_Y, V_{XZ})=(100,100)$\;km/s.
Surprisingly, however, there are many stars with thick disk-like metallicities found in the halo region of the Toomre diagram, and some of them are on very retrograde orbits.

\begin{figure}
\begin{center}
\includegraphics[width=0.9\columnwidth]{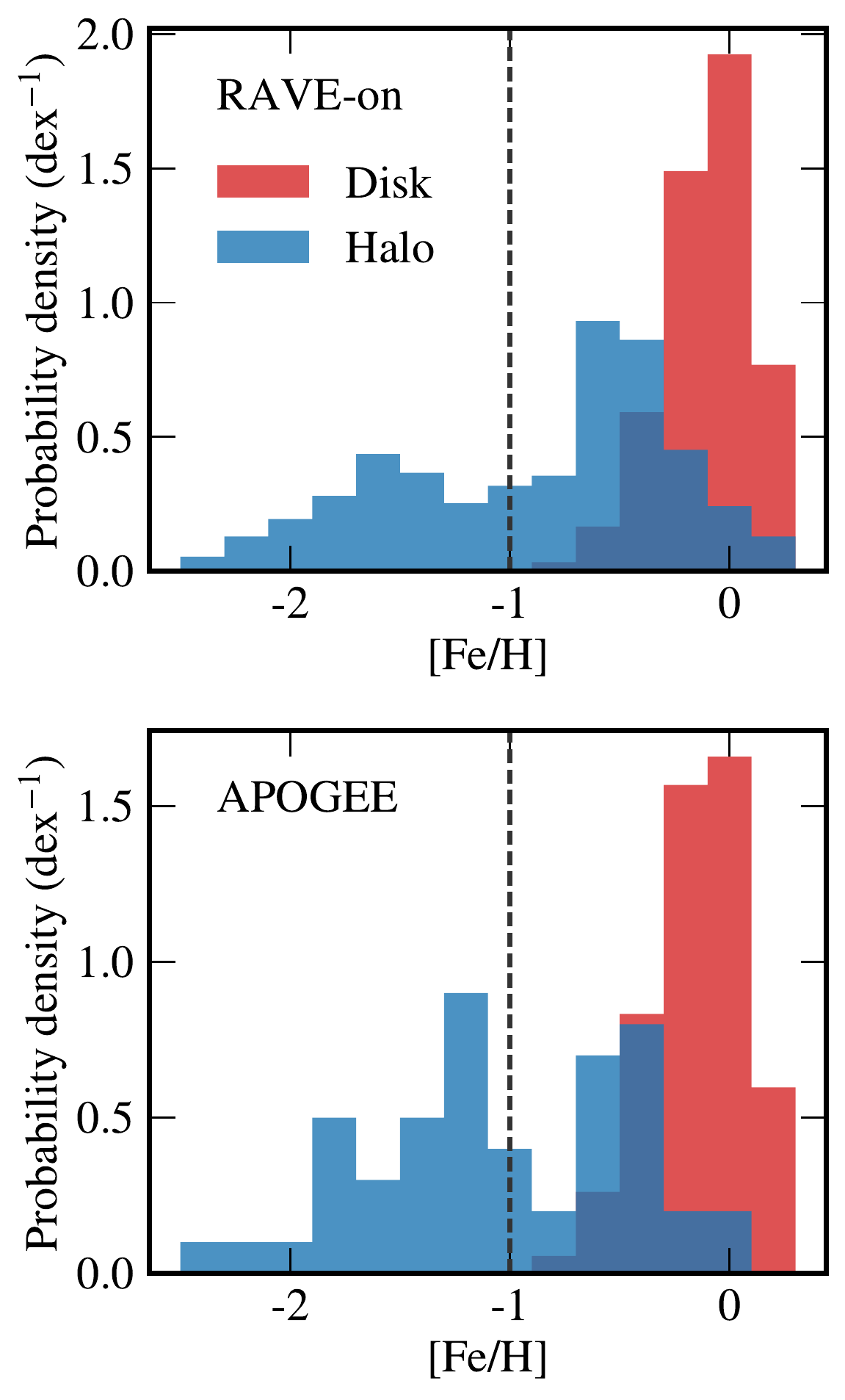}
\caption{Metallicity distribution function of the Solar neighborhood in TGAS and RAVE-on catalogs on the top, and TGAS and APOGEE at the bottom.
Kinematically-selected disk stars are shown in red, while the halo distribution is plotted in blue.
In both samples, there is a population of metal-rich halo stars, with $\approx50\%$ of stars having $\rm[Fe/H]>-1$ (marked with a vertical dashed line).}
\label{fig:mdf}
\end{center}
\end{figure}

Figure~\ref{fig:mdf} (top) shows the metallicity distribution for the two kinematic components identified above: the disk in red and the halo in blue.
The disk is more metal rich than the halo, and peaks at the approximately solar metallicity, $\rm[Fe/H]=0$.
The halo is more metal poor, and exhibits a peak at $\rm[Fe/H]\sim-1.6$, typical of the inner halo \citep[e.g.,][]{allende-prieto2006}.
However, the metallicity distribution of the halo has an additional peak at the metal-rich end, centered on $\rm[Fe/H]\sim-0.5$ and extending out to the super-solar values.

To corroborate the existence of metal-rich stars on halo orbits, we also show the metallicity distribution function for TGAS stars observed with APOGEE at the bottom of Figure~\ref{fig:mdf}.
The disk--halo decomposition for the APOGEE sample was performed in the identical manner to that of RAVE-on.
The metallicity distributions between the two surveys are similar: the disk stars are metal-rich, while the halo has a wide distribution ranging from $\rm[Fe/H]\sim-2.5$ to $\rm[Fe/H]\sim0$.
A bimodality is present in the metallicity distribution of APOGEE halo stars, although it is less prominent than in the RAVE-on sample due to the smaller sample size.
The apparent bimodality in the metallicity distribution of RAVE-on halo stars is slightly more metal-poor, $\rm[Fe/H]\approx-1.1$, than observed in the APOGEE sample, $\rm[Fe/H]\approx-0.8$.
In what follows, we compromise between these two values, and split the halo sample at $\rm[Fe/H]=-1$, into a metal-rich ($\rm[Fe/H]>-1$) and a metal-poor component ($\rm[Fe/H]\leq-1$).

\begin{figure}
\begin{center}
\includegraphics[width=0.9\columnwidth]{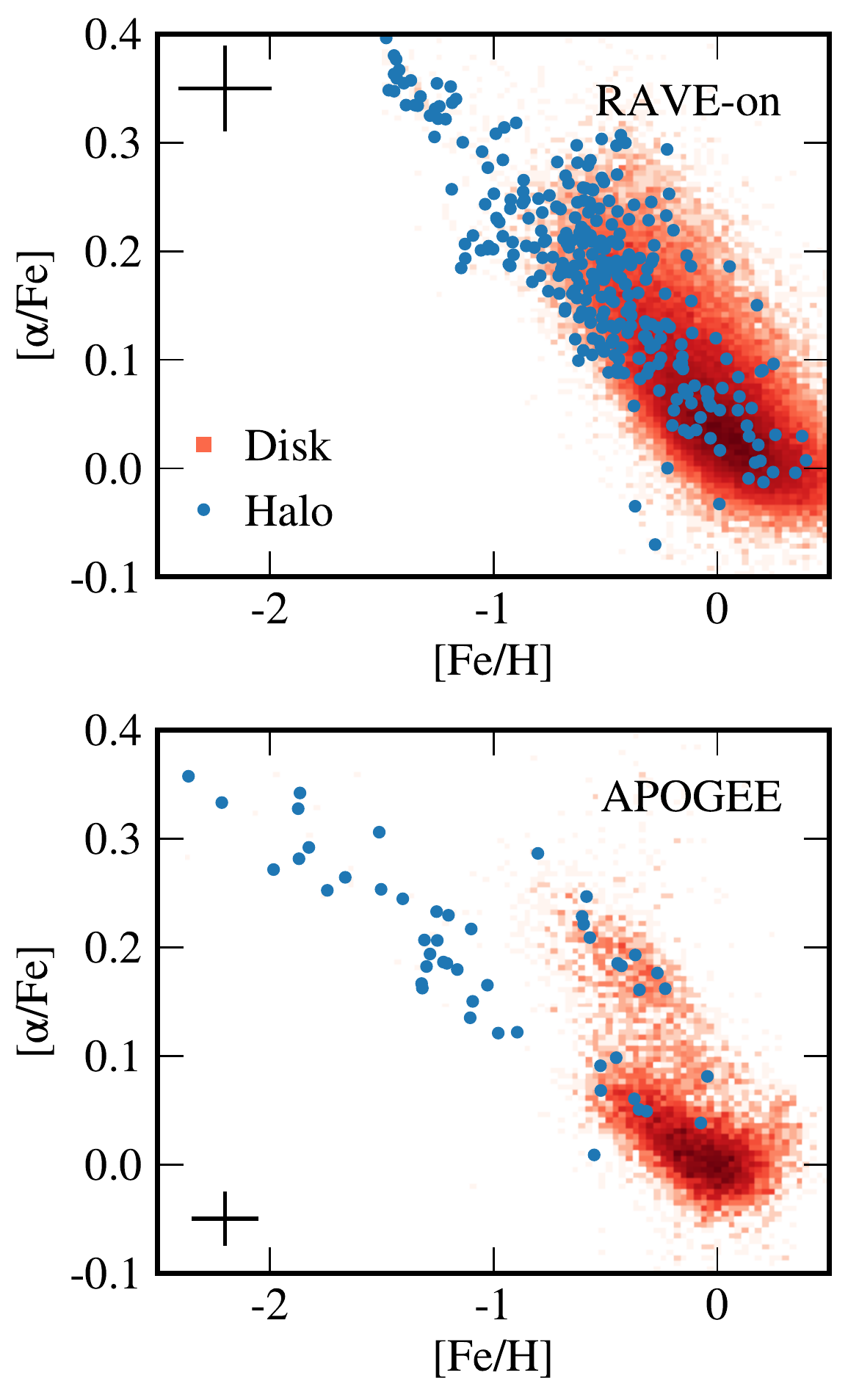}
\caption{Chemical abundance pattern, [$\alpha$/Fe] vs [Fe/H], for TGAS--RAVE-on sample on the top and TGAS--APOGEE at the bottom.
The pattern for disk stars is shown as a red-colored Hess diagram (logarithmically stretched), while the halo stars are shown individually as blue points.
In both surveys, the metal-poor halo is $\alpha$-enhanced, while the metal-rich halo follows the abundance pattern of the disk.}
\label{fig:afeh}
\end{center}
\end{figure}

Chemical abundances have been used to discern different components of the Galaxy \citep[e.g.,][]{gilmore1989}.
The abundance space of [$\alpha$/Fe] vs [Fe/H] is particularly useful in tracing the origin of individual stars \citep[e.g.,][]{lee2015}.
Figure~\ref{fig:afeh} shows this space for RAVE-on spectra on the top, and APOGEE on the bottom.
The disk distribution is shown as a red density map, while the less numerous halo stars are shown individually in blue.
Similarly to the overall metallicity distribution function, RAVE-on and APOGEE surveys are in a qualitative agreement in terms of the more detailed chemical abundance patterns as well.
At low metallicities, the halo is $\alpha$-enhanced, but at high metallicities its [$\alpha$/Fe] declines, following the disk abundance pattern, both in terms of the mean [$\alpha$/Fe] and its range at a fixed [Fe/H].
In particular, thick disk stars have higher $\alpha$-element abundances at a given metallicity \citep[e.g.,][]{nidever2014}, and follow a separate sequence visible in the more precise APOGEE data ([Fe/H]$\sim-0.2$, [$\alpha$/Fe]$\sim0.2$), while the thin disk is in general more metal-rich and $\alpha$-poor.
Metal-rich halo stars in both samples span this range of high- and low-$\alpha$ abundances.
At lower metallicities, \citet{ns2010} reported that halo stars follow two separate [$\alpha$/Fe] sequences, with the high-$\alpha$ stars being on predominantly prograde orbits, whereas the low-$\alpha$ stars are mostly retrograde.
We do not resolve the two sequences, or see any correlation between the orbital properties and [$\alpha$/Fe] at a fixed [Fe/H].
However, the RAVE-on abundances are fairly uncertain (typical uncertainty is $\sim0.07$\;dex, marked by a black cross on the top left in Figure~\ref{fig:afeh}), so the sequences seen in the higher-resolution data from \citet{ns2010} would not be resolved in this data set.
In the APOGEE sample, where the typical uncertainty is smaller, $\sim0.03$\;dex, the number of halo stars is too small to unambiguously identify multiple $[\alpha/\rm Fe]$ sequences.

The evidence presented so far shows that the stellar halo in the Solar neighborhood has a distinct metal-poor and a metal-rich component.
Given that the metal-rich component follows the abundance pattern of the disk, we discuss the possible contamination by the thick disk in Appendix~\ref{sec:tdcontamination}, and rule out the possibility that a sizeable fraction of the metal-rich halo is attributable to the canonical thick disk.
In the next section, we proceed to characterize the orbital properties of the two halo components.

\begin{figure}
\begin{center}
\includegraphics[width=\columnwidth]{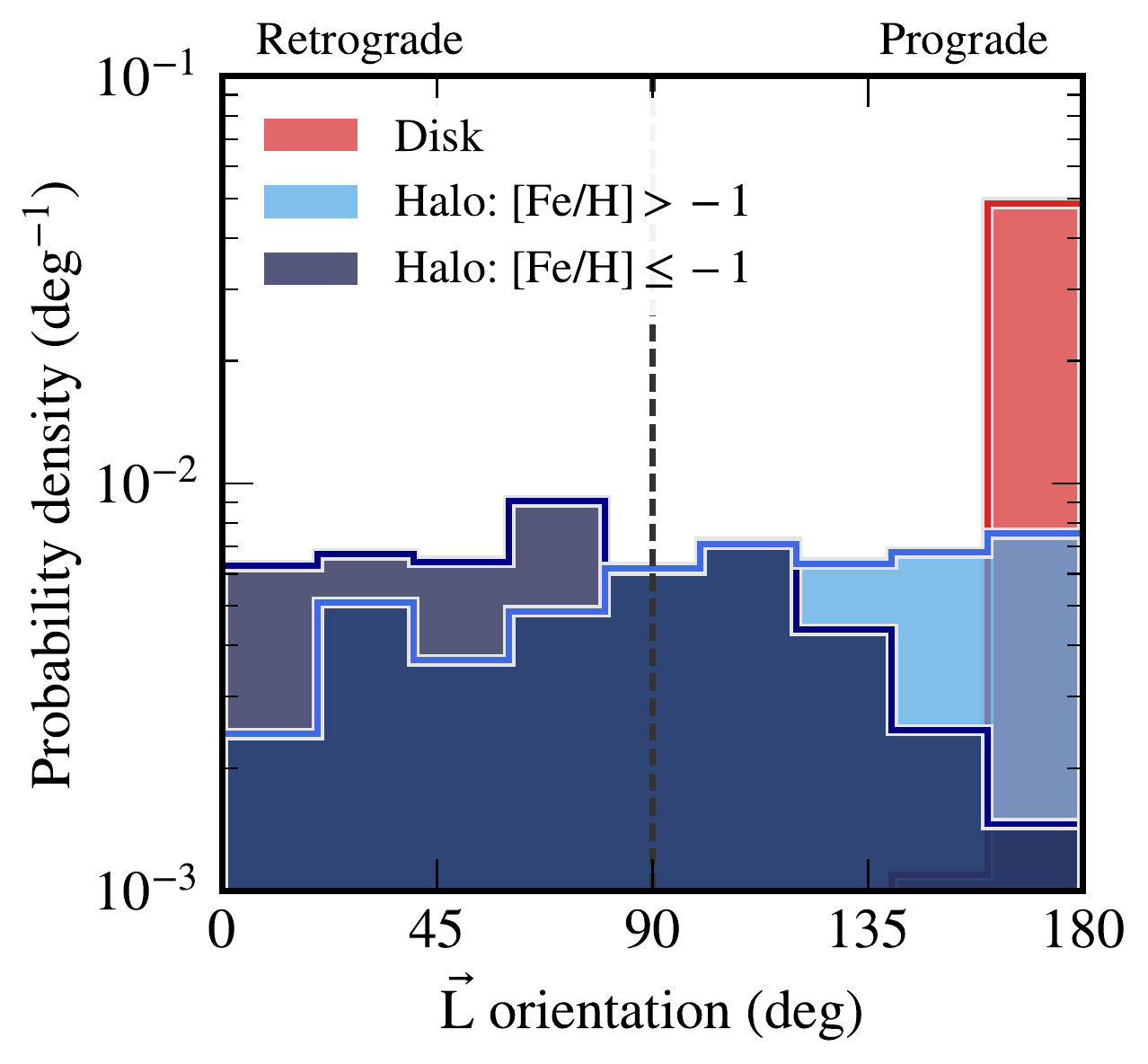}
\caption{Orientation of angular momenta with respect to the Galactocentric $Z$ axis for different Galactic components: the disk in red, metal-poor halo in dark blue and metal-rich halo in light blue.
The angular momenta of disk stars are aligned with the $Z$ axis, while those of halo stars are more uniformly distributed.
There is an excess of metal-rich halo stars on prograde orbits, $\theta_L>90^\circ$, compared to the metal-poor halo orbital orientations.}
\label{fig:ltheta}
\end{center}
\end{figure}

\subsection{Angular momenta}
\label{sec:l}
Stellar orbits can be classified in terms of their integrals of motion, however, calculating these requires the knowledge of the underlying gravitational potential \citep{bt2008}.
Furthermore, in a realistic Galactic environment, some stars are on chaotic orbits \citep[e.g.,][]{price-whelan2016}, where the integrals of motion do not exist.
On the other hand, any star with a measured position in a 6D phase space has a well defined angular momentum.
In this section we use angular momenta as empirical diagnostics of stellar orbits, and focus in particular on the orientation of the angular momentum vector with respect to the Galactocentric $Z$ axis, quantified by the angle
\begin{equation}
\theta_L \equiv \arctan(L_Z/|\vec{L}|)
\label{eq:thetal}
\end{equation}
where $L_Z$ is the $Z$ component of the angular momentum, and $|\vec{L}|$ its magnitude.
$L_Z$, and hence $\theta_L$, are conserved quantities in static, axisymmetric potentials, such as that of the Milky Way disk.
This has already been utilized to identify coherent structures in the phase space of local halo stars \citep[e.g.,][]{helmi1999, smith2009}.
In the adopted coordinate system, the disk orientation angle is $\theta_L=180^\circ$, so that prograde orbits are those with $\theta_L>90^\circ$ and retrograde have $\theta_L<90^\circ$.
We show the distribution of angular momentum orientations $\theta_L$ for the identified Galactic components in Figure~\ref{fig:ltheta}.

\begin{figure}
\begin{center}
\includegraphics[width=0.9\columnwidth]{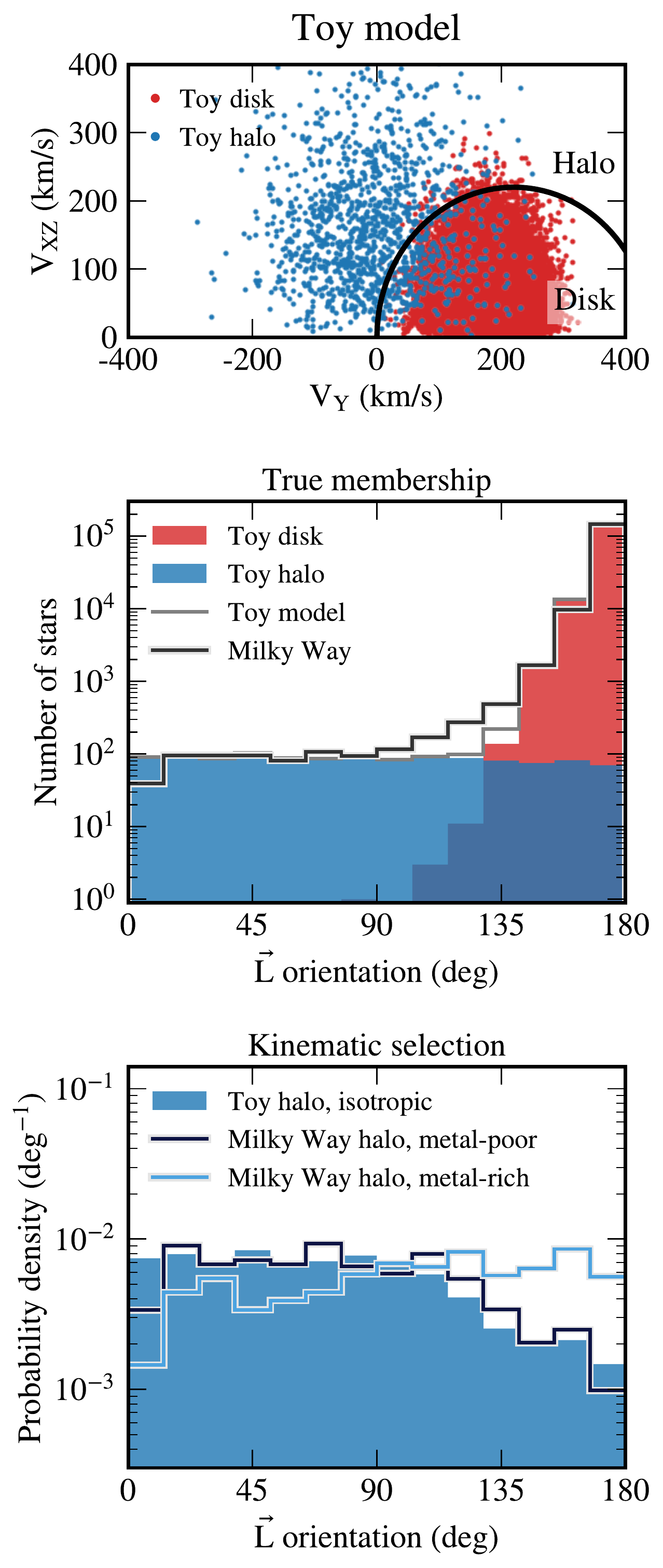}
\caption{Toy model for the phase space of the solar neighborhood.
The model consists of a halo (blue) and a disk component (red), with their positions drawn directly from the TGAS--RAVE-on sample, and kinematics from the velocity ellipsoids of \citet{bensby2003}.
Top panel shows the model in the Toomre diagram.
The black line is the employed demarcation between the halo and the disk, which does fairly good job in separating the two in the toy model as well.
The middle panel shows the orientation of angular momenta in the model, with each component shown as a shaded histogram, and a model total with a black line.
Toy model angular momenta successfully reproduce the angular momentum orientations observed in the Milky Way (gray dashed line).
Kinematically selecting the halo in the model (bottom panel, shaded histogram) produces a distribution in excellent agreement with the distribution of metal-poor halo stars in the Milky Way (dark blue line).
The metal-rich halo in the Milky Way (light blue line) is inconsistent with being a part of an isotropic halo studied in this toy model.}
\label{fig:toy}
\end{center}
\end{figure}

As expected, most of the disk stars are indeed on orbits in the disk plane with $V_Z\approx0$, and have $\theta_L\approx180^\circ$ (red histogram in Figure~\ref{fig:ltheta}).
Angular momenta of both halo components span the entire range of $0^\circ<\theta_L<180^\circ$, but in detail, their distributions are significantly different from each other.
The metal-poor halo has almost a flat distribution as a function of $\theta_L$, with a slight depression at very prograde angles (dark blue histogram in Figure~\ref{fig:ltheta}).
The metal-rich halo is predominantly prograde, but also has a long tail to retrograde orbits (light blue histogram in Figure~\ref{fig:ltheta}).
In the next section, we explore the origin of these distributions in terms of a toy model for the kinematic distribution of the Galaxy, as well as in comparison to a hydrodynamical simulation of a Milky Way-like galaxy.

\section{Origin of halo stars in the Solar neighborhood}
\label{sec:origin}
In the previous section, we presented a metal-rich component of the stellar halo in the solar neighborhood that is preferentially on prograde orbits with respect to the Galactic disk.
In this section, we test whether the observed orientations of these angular momenta can be understood in the context of a simple toy model (\S\ref{sec:toymodel}), as well as in comparison to a solar-like neighborhood in a cosmological hydrodynamical simulation (\S\ref{sec:latte}).
Finally, we study the formation paths of the simulated stellar halo to provide a possible origin for this metal-rich halo component in \S\ref{sec:formation}.

\subsection{Toy model}
\label{sec:toymodel}
To construct a toy model for our TGAS--RAVE-on sample, we assign stars to one of the three main components of the Milky Way galaxy: a thin disk, a thick disk, or a halo \citep[e.g.,][]{bhg2016}.
The model is defined by the number of stars in each component, their spatial and kinematic properties.
To set the number of halo stars in the toy model, we assumed that the halo is isotropic.
In that case, an equal number of halo stars are on prograde and retrograde orbits. 
Since we expect no contamination from the disk on retrograde orbits, we set the total size of the halo in the toy model to be twice the number of retrograde stars in our Milky Way sample.
For the remaining disk stars, we vary the ratio of thin to thick disk stars to best match the distribution of prograde orbits.

Once the number of stars in each component had been determined, we proceeded to assign them their phase space coordinates.
Our sample is spatially confined to within only several kpc from the Sun (see Appendix~\ref{sec:distances}), so we see no differences in spatial distributions of the kinematically defined components from section \S\ref{sample}.
This allowed us to take the spatial distribution of stars in our sample, and randomly designate them a component in the toy model, thus ensuring that the spatial selection function of both TGAS and RAVE-on is properly reproduced in the model.
For each star in the model, we drew a 3D velocity from its component's velocity ellipsoid measured by \citet{bensby2003} on a smaller sample of local stars with Hipparcos parallaxes and proper motions, and which accounts for the asymmetric drift.
With positions and velocities in place, we calculated the angular momenta and their orientation angles with respect to the $Z$ axis, $\theta_L$, for all stars in the toy model.

The properties of our toy model are summarized in Figure~\ref{fig:toy}.
The top panel shows the components of the model in the Toomre diagram, with the disk stars in red, halo stars in blue, and the black line delineating our kinematic boundary between the halo and the disk.
The velocity ellipsoids of these components overlap, so our kinematic definition of a halo produces a sample which is likely neither pure, nor complete.
This is illustrated in the toy model by several thick disk stars that enter the halo selection box at $(V_Y, V_{XZ}) \simeq (100,200)$\;km/s, and also halo stars on prograde orbits, which fall outside of the halo selection.
As no simple kinematic cut will completely separate the different components, we opted to emphasize the purity of our halo sample.
Based on the toy model, we estimate that the fraction of interloping disk stars in a kinematically defined halo is only $\sim10\%$, but this in turn makes the halo sample less complete at $75\%$.
For a comparison, \citet{ns2010} defined a halo that is $\sim90\%$ complete, but only $\sim55\%$ pure.

\begin{figure}
\begin{center}
\includegraphics[width=0.9\columnwidth]{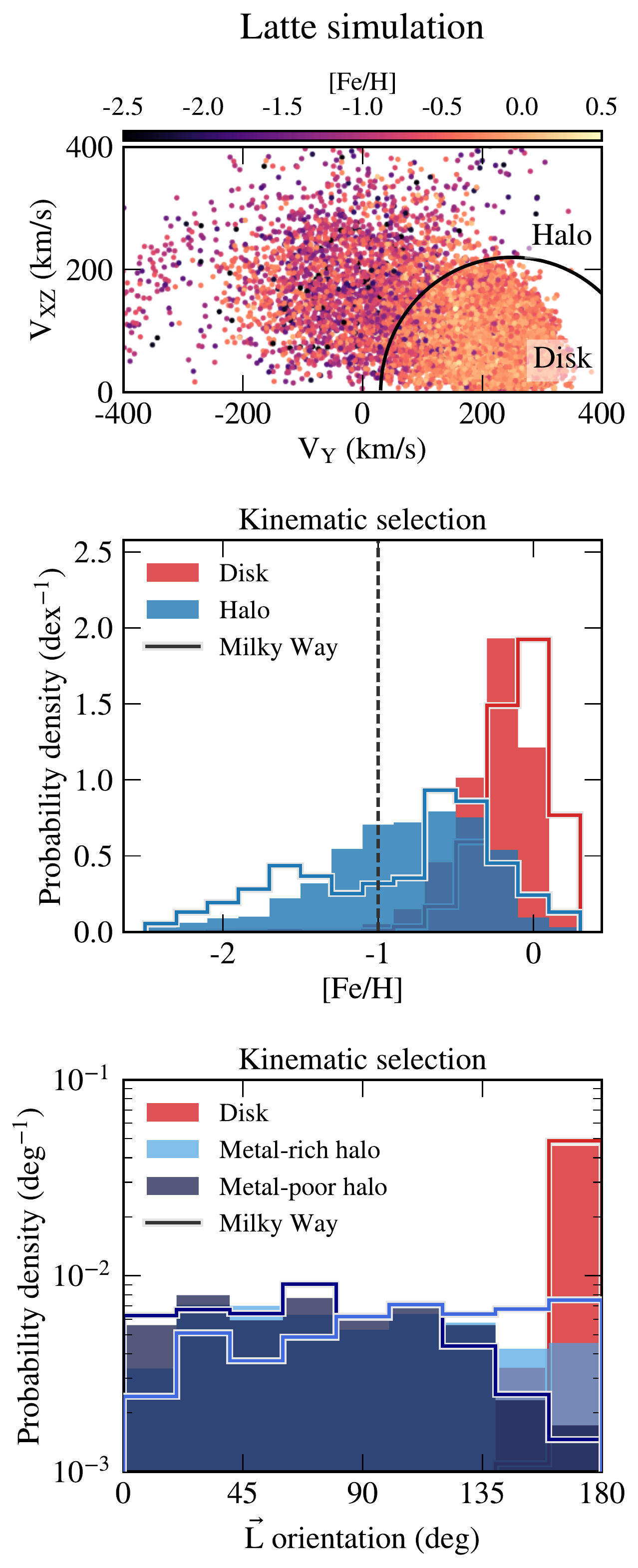}
\caption{Stars from the Latte cosmological zoom-in baryonic simulation of a Milky Way-mass galaxy, from the FIRE project, observed identically to stars in the Solar neighborhood.
Top panel shows positions of stars in the Toomre diagram, color-coded by metallicity.
We identify disk and halo stars in Latte using a cut in this diagram (black line), analogous to a cut used for the Milky Way sample.
Metallicity distributions of Latte stars are in the middle, with disk being shaded red, and halo blue.
Bottom panel shows the orientations of stars' angular momenta in Latte, with disk in red, metal-poor halo in dark blue, and metal-rich halo in light blue.
Properties of Latte stars are remarkably similar to the distributions of stars in our Milky Way sample (reproduced as empty histograms on the middle and bottom panel for direct comparison).
}
\label{fig:latte}
\end{center}
\end{figure}

The middle panel of Figure~\ref{fig:toy} shows the orientation of angular momenta, $\theta_L$, for the components of the toy model, with halo in blue and disk in red.
As expected, the majority of disk stars are moving in the disk plane, and the distribution is sharply peaked at $\theta_L=180^\circ$.
The nearly isotropic halo of the \citet{bensby2003} model has a flat distribution in $\theta_L$.
The sum of the two components is represented with a thick black line, which compares favorably to the distribution of $\theta_L$ observed in the Milky Way, and shown in dashed gray.
The agreement between the toy model and the Milky Way is particularly good at very prograde and very retrograde orbits.
The abundance of stars on only slightly prograde orbits ($\theta_L\approx100^\circ$) is somewhat underestimated in the toy model, indicating that the transition between a disk and a halo component in the Milky Way is more gradual than what can be reproduced by a simple model featuring only two disks and an isotropic halo.

The bottom panel of Figure~\ref{fig:toy} compares the modeled halo and the observed one.
For a fair comparison, in this panel we only consider model stars that would satisfy our kinematic halo selection, which are a combination of some disk and the majority of halo stars, as discussed above.
The distribution of $\theta_L$ for this kinematically-selected halo stars from a toy model is shown as a shaded histogram.
While the intrinsic distribution of an isotropic halo is flat (middle panel), kinematic selection introduces a suppression at the most prograde orbits (right panel).
Kinematic selection excludes halo stars with $|V-V_{LSR}|\leq220$\;km/s, all of which are on prograde orbits, which is manifested as a depression at $\theta_L\gtrsim90^\circ$.
The magnitude of this depression exactly matches the distribution of metal-poor halo stars in the Milky Way, overplotted with a dark blue line in the right panel of Figure~\ref{fig:toy}.
This suggests that the metal-poor halo in the Solar neighborhood is intrinsically isotropic, and that at least some stars more metal-poor than $\rm[Fe/H]\leq-1$, which are kinematically consistent with the disk, are in fact a part of the metal-poor halo.
The distribution of metal-rich halo stars in the Milky Way, shown as light blue line in the bottom panel of Figure~\ref{fig:toy}, shows the opposite behavior of an excess at prograde orbits and is significantly different from the toy model prediction for an isotropic halo.

A simple toy model successfully explains the bulk properties of our sample: most of the stars are in a rotating disk, with a minority in an isotropic halo, which maps well to the metal-poor halo stars identified in our TGAS--RAVE-on sample.
This toy model also points out that the metal-rich halo stars are inconsistent with either of its components.
Next, we analyze the origin of such a population using a hydrodynamical simulation.

\subsection{The Latte simulation}
\label{sec:latte}
The Latte simulation, first presented in \citet{wetzel2016}, is a simulation from the Feedback In Realistic Environments (FIRE)\footnote{The FIRE project website is \url{http://fire.northwestern.edu}} project \citep{hopkins2014FIRE}. Latte is fully cosmological, with baryonic mass resolution of $7000 \,M_{\sun}$ and spatial softening/smoothing of $1\,pc$ for gas and $4\,pc$ for stars.
The simulation uses the standard FIRE-2 implementation of gas cooling, star formation, stellar feedback, and metal enrichment as described in \citet{hopkins2017}, including the numerical terms for turbulent diffusion of metals in gas as described therein\footnote{
The original simulation presented in \citet{wetzel2016} did not include terms for turbulent metal diffusion.
Here, we analyze a re-siumulation that includes those terms (all other parameters unchanged), which creates a more realistic metallicity distribution function in both the host galaxy (Wetzel et al., in prep.) and its satellites (Escala et al., in prep.).
As explored in \citet{Su2016} and \citet{hopkins2017}, the inclusion of turbulent metal diffusion has no systematic effect in any gross galaxy properties (including the average metallicity), as we also checked for all analyses in this paper.
}, and the GIZMO code \citep{hopkins2015}. FIRE simulations have been used to study galaxies from ultra-faint dwarf galaxies \citep{wheeler2015} to small groups of galaxies \citep{feldmann2016}; FIRE simulations successfully reproduce observed internal properties of dwarf galaxies \citep{chan:fire.dwarf.cusps, elbadry2016}, thin/thick disk structure and both radial and vertical disk metallicity gradients in a Milky Way-like galaxy \citep{ma2016}, star-formation histories of dwarf and massive galaxies \citep{hopkins2014, sparre:sfmainsequence}, global galaxy scaling relations \citep{hopkins2014, ma:mass.metallicity, feldmann2016}, and satellite populations around Milky Way-mass galaxies \citep{wetzel2016}.

To construct a sample of star particles in Latte analogous to the TGAS--RAVE-on sample, we first aligned the simulation coordinate system with the disk, and then selected stars in a 3\;kpc sphere located at a distance of 8.3\;kpc from the center of the galaxy.
We classified star particles as either disk or halo using a kinematic cut in the Toomre diagram similar to the one employed for stars in the Milky Way (\S\ref{sec:sample}).
Because the circular velocity in Latte is slightly different from the Galactic value, the definition of the local standard of rest is also different, but we keep the same conservative measure for the dispersion of 220\;km/s in the disk of Latte.
The Toomre diagram for the sample in Latte is shown in the top panel of Figure~\ref{fig:latte}.
Qualitatively, it is similar to that of the Milky Way (Figure~\ref{fig:toomre}), with most of the stars rotating in the disk at $\sim235$\;km/s, and the density of stars smoothly decreasing away from the local standard of rest.
Quantitatively, the halo fraction in Latte is an order of magnitude higher at 10\%, and, compared to the Milky Way's, its kinematic space is more structured.
This is partly because the Latte sample does not suffer from selection effects, so it effectively extends to larger distances, where the halo constitutes a larger mass fraction.
Any kinematic structures are hence better sampled and more readily observable in Latte.
Additionally, at least some of the structure present in the Milky Way sample is smoothed by the observational uncertainties \citep[see, e.g.,][]{sanderson2015}, which are not present in Latte.

Stars in the Toomre diagram of Figure~\ref{fig:latte} are color-coded by metallicity and show trends similar to those observed in the Milky Way.
For a more quantitative analysis, we show metallicity distribution of Latte's disk and halo components in the central panel of Figure~\ref{fig:latte}.
To facilitate comparison, we also include metallicity distributions of the Milky Way components as empty histograms.
The Latte halo is more metal poor than its disk, and although there is no bimodality in the halo metallicity, the whole distribution is as wide as observed in the Milky Way, extending from [Fe/H]$\lesssim-2$ to [Fe/H]$\simeq0$.
This agreement, in addition to $\rm[\alpha/Fe]$ abundance trends recovered in simulated disks (Wetzel et al., in prep), demonstrate that the physics included in the FIRE simulations captures the most important ingredients for chemical evolution in galaxies.
Following our analysis of the Milky Way sample, we proceed to divide the Latte halo into a metal-rich, [Fe/H]$>-1$, and a metal-poor, [Fe/H]$\leq-1$, component.
The ratio of metal-rich to metal-poor halo stars in Latte is not quite the same as in the Milky Way, but this does not seriously impede our goal of qualitatively understanding differences between the two populations.

Finally, we analyze orbital properties of stars in Latte by showing the orientation of their angular momenta with respect to the $Z$ axis, $\theta_L$, as solid histograms in the bottom panel of Figure~\ref{fig:latte}.
The angular momenta of Latte disk stars are well aligned with the $Z$ axis, with $\theta_L\simeq180^\circ$, similar to disk stars in the Milky Way (shown in Figure~\ref{fig:latte} as an empty histogram).
The Latte halo shows a flatter distribution of $\theta_L$, but there is still an excess of metal-rich stars on prograde orbits (histogram shaded light blue) with respect to the metal-poor halo stars (dark blue).
Overall, Latte stars have similar kinematic, chemical and orbital properties to stars observed in the Milky Way, suggesting that it provides a reasonable analogue.
We next trace Latte stars back to their birth location and suggest a possible scenario for the formation of halo stars in the Solar neighborhood.

\subsection{Formation paths of the Latte stellar halo}
\label{sec:formation}
\begin{figure}
\begin{center}
\includegraphics[width=0.9\columnwidth]{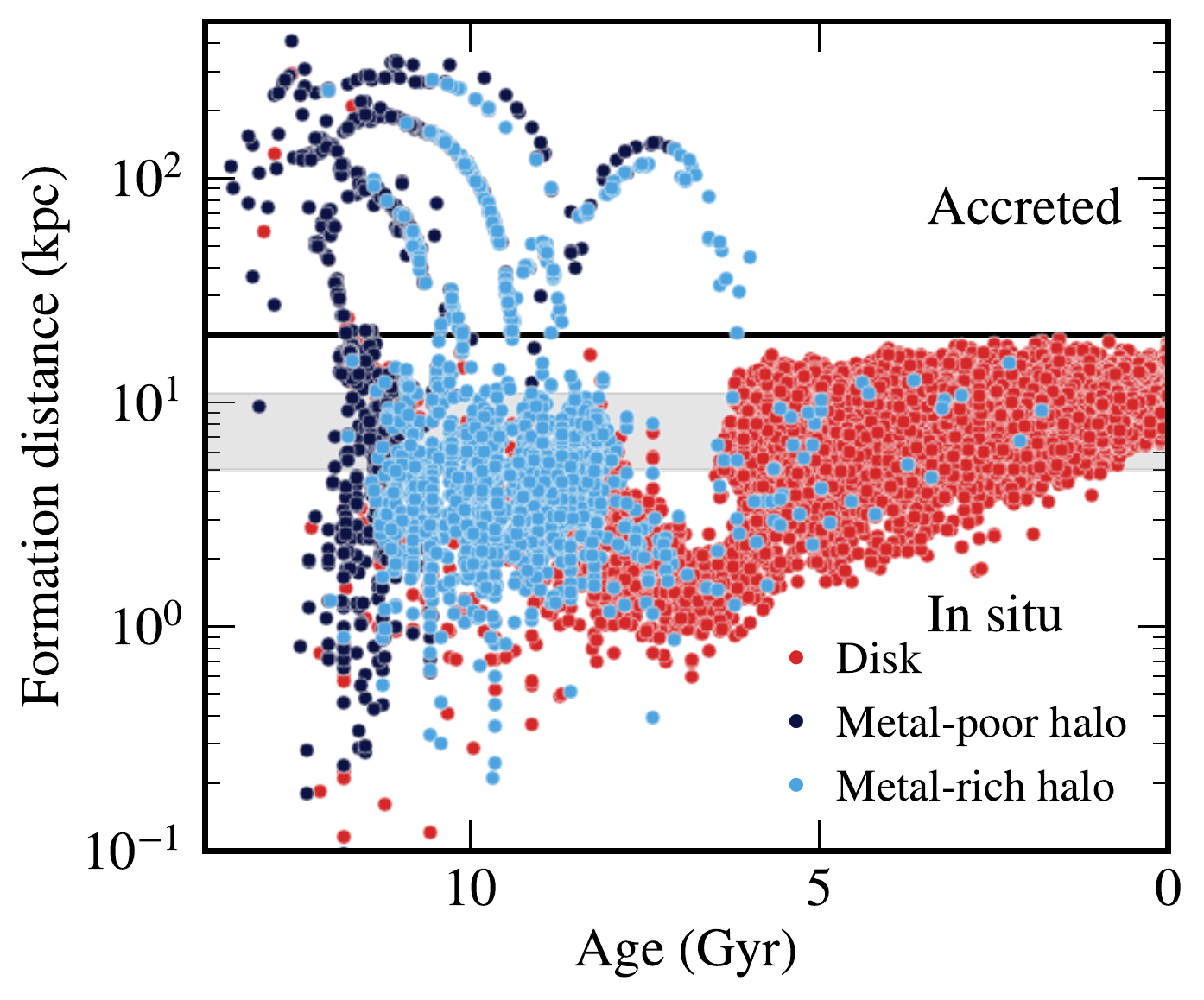}
\caption{Distance from the host galaxy at the time of formation, as a function of current stellar age (in physical kpc), for star particles currently in the solar neighborhood (gray shaded region) in the Latte simulation.
Blue circles are kinematically identified as halo (with metal-rich stars plotted in light blue, and metal-poor are in dark blue), red circles are kinematically consistent with the disk.
The galaxy experienced a significant merger 7\;Gyr ago, so most stars formed at that time come from the inner 1\;kpc.
We define a star as being accreted if it formed more than 20\;kpc from the host galaxy (horizontal black line).
Otherwise, we define its formation as in situ (in the disk).
Current disk stars formed in situ and at later times, while halo stars formed early.
Furthermore, the halo contains a population of both accreted and in situ stars, with the accreted component being more metal-poor.}
\label{fig:dform}
\end{center}
\end{figure}

We define the formation distance of a star particle in Latte as its physical (not comoving) distance from the center of the primary (Milky Way-like) galaxy at time that it formed, and we inspect this formation distance as a function of the star's current ($z = 0$) age in Figure~\ref{fig:dform}.
Stars with disk kinematics (as defined by our Toomre-diagram cut in Figure~\ref{fig:latte}) are shown in red, while those identified as halo are blue.
The suppression of stars that formed at solar distances $\sim$7\;Gyr ago coincides with the time of the last significant merger ($z\sim0.7$).
This event brought significant gas to the center of the galaxy, which switched the star-forming conditions from those producing mainly stars on halo-like orbits (more than 95\% of halo stars are older than 6\;Gyr) to the orderly production of disk stars (more than 90\% of disk stars were formed in the last 6\;Gyr).
The formation distances between the disk and the halo component are equally dichotomous: most of the disk stars were formed close to their present-day distance from the galactic center ($5-11$\;kpc, shaded gray in Figure~\ref{fig:dform}), while halo stars originated from the extremes of the central 1\;kpc to the virial radius and beyond (for reference, the virial radius 10\;Gyr ago was $\approx70\;$kpc).
Overall, only $\sim20$\% of Latte stars on halo orbits were formed inside of their present-day radial distance range, indicating that radial redistribution is an important phenomenon sculpting the inner Galaxy.
We discuss the implications of the halo component undergoing radial migration/heating in \S\ref{sec:migration}.

The formation distance of a star particle can also be used as a rudimentary diagnostic of its formation mechanism.
From Figure~\ref{fig:dform} we see that almost none of the disk stars were formed beyond 20\;kpc (indicated with a horizontal black line), so we adopt this distance as a delimiter between the in-situ and accreted origin for stars in Latte.
Most of the accreted stars are classified as halo, and were formed inside dwarf galaxies merging with the host galaxy.
The tracks in the space of formation distance and age (Figure~\ref{fig:dform}) delineate the orbits of these dwarf galaxies.
All of the satellites are disrupted once they get within the central 20\;kpc, and for most of them this happens on the first approach.
In the process, they bring in gas which fuels in-situ star formation.
At early times, most of the stars that formed in situ also become a part of the halo, while the last significant merger starts the onset of the in-situ formation of the disk.
Although satellite accretion onto the host galaxy continues after the last significant merger, none of the accreted stars reach the solar circle, so effectively all of the halo stars in the solar neighborhood were formed prior to the last significant merger.
This is in line with findings of \citet{zolotov2009}, who showed that late-time accretion predominantly builds outer parts of the stellar halo.

In summary, Latte stars born at late times formed in the disk, while old stars in Latte are predominantly part of the halo.
One third of the halo in the solar neighborhood was accreted from the infalling satellites, while the majority of stars formed in situ in the inner galaxy and migrated to the solar circle.
In this simplified depiction for the origin of the stellar halo, we do not account for the unlikely possibility that stars that formed within 20\;kpc still could have been bound to a satellite galaxy, and hence accreted, nor do we distinguish between different modes of in-situ halo formation (e.g., through a dissipative collapse; \citealt{samland2003}, or with stars being heated from the disk; \citealt{purcell2010}).
These are not serious shortcomings, because satellites that entered the inner 20\;kpc were disrupted soon thereafter, so any newly-formed stars would have been only loosely bound to the satellite at the time.
Furthermore, given that the median formation distance of the in-situ halo is $\sim4$\;kpc, we expect that dissipative collapse only marginally contributes to the census of halo stars at the solar circle, but ultimately we draw conclusions that are insensitive to the particulars of their origin.

\begin{figure*}
\begin{center}
\includegraphics[width=\textwidth]{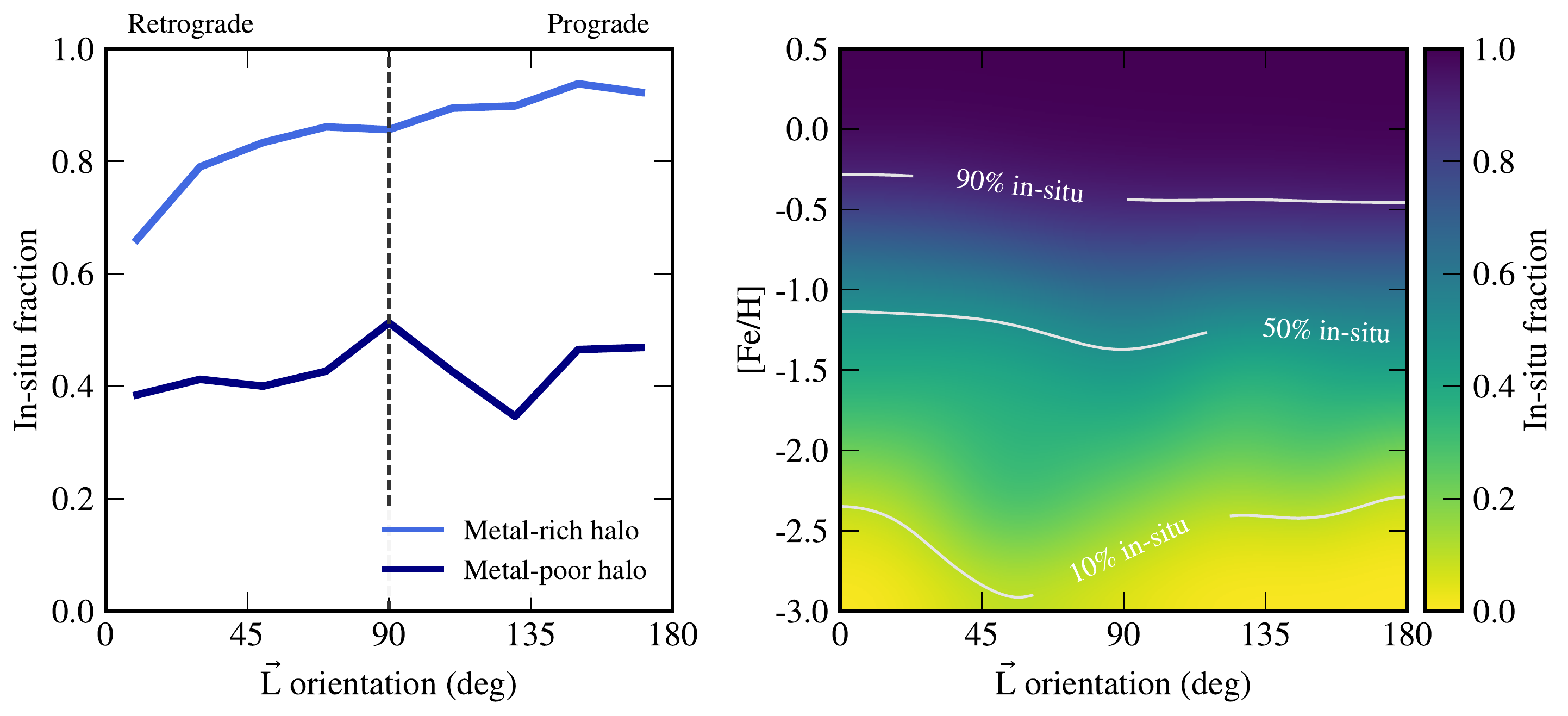}
\caption{(\emph{Left}) Average in-situ fraction of Latte halo stars in the solar neighborhood as a function of angular momentum orientation.
Metal-poor halo stars (dark blue) have a $\approx40\%$ probability of being formed in situ regardless of their angular momentum.
The metal-rich halo (light blue) is more likely to have formed in situ, and this probability increases for stars on orbits aligned with the disk ($\theta_L>90^\circ$).
(\emph{Right}) Average in-situ fraction as a function of metallicity and current angular momentum orientation angle.
Whether or not a star was accreted depends only weakly on its current orbital properties, but it correlates well with its metallicity.
The in-situ fraction varies smoothly between the metal-rich end, where all stars formed in situ, and the metal-poor end, where all stars were accreted.
Thus, in the inner halo, a star's metallicity is a much better indicator of its origin (the probability that it was formed in situ) than its current kinematics.
}
\label{fig:facc}
\end{center}
\end{figure*}

In Figure~\ref{fig:facc} we explore how the fraction of halo stars formed in situ depends on the angular momentum orientation ($\theta_L$, left) and a combination of angular momentum and metallicity (right).
On average, 40\% of metal-poor halo stars have formed in situ (dark blue line in the left panel of Figure~\ref{fig:facc}).
Metal-rich halo stars are more likely to have formed in situ (light blue line), and this probability increases to 90\% for stars whose orbits are aligned with the disk.
Interestingly, when we investigate the dependence of the in-situ fraction simultaneously as a function of metallicity and angular momentum orientation (right panel of Figure~\ref{fig:facc}), we find only a weak dependence on the current angular momentum orientation of a star, but a strong correlation with metallicity.
All of the lowest metallicity stars, [Fe/H]$\lesssim-2.5$, were accreted (yellow), and all of the metal-rich halo stars, [Fe/H]$\gtrsim-0.5$, formed in situ (purple).
Given the similarities between the global chemical and orbital properties in Latte and the Milky Way, this result suggests that the metal-rich halo component identified in the TGAS--RAVE-on sample formed in the inner Galaxy, but was driven to the Solar circle through subsequent evolution.

\begin{figure}
\begin{center}
\includegraphics[width=0.9\columnwidth]{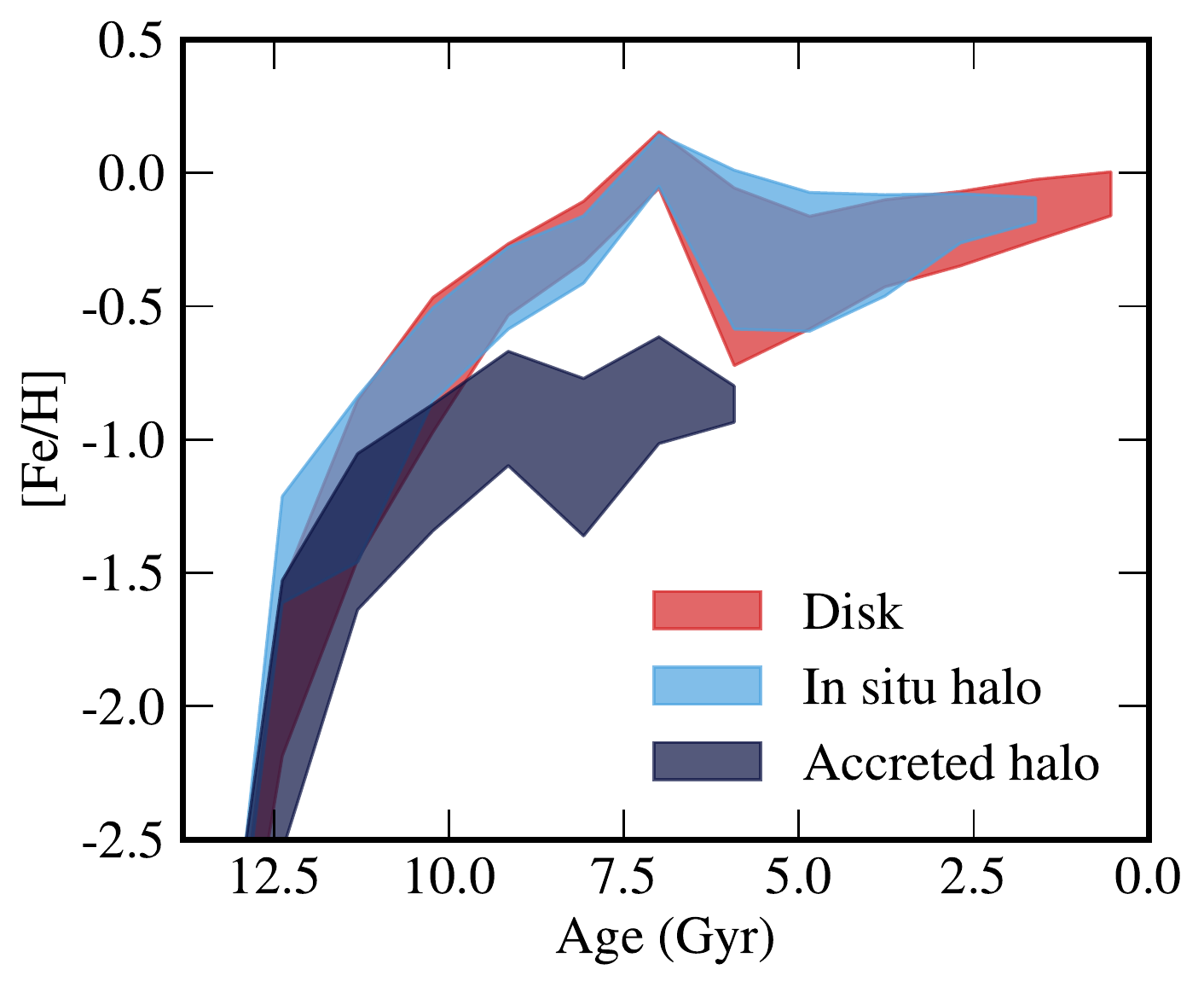}
\caption{Metallicity range (16--84 percentile) of star particles in the Latte simulation as a function of their age for three structural components (defined as in Figures~\ref{fig:latte}--\ref{fig:dform}) identified at the solar circle: disk (red), in-situ halo (light blue) and accreted halo (dark blue).
The in-situ halo follows the metallicity evolution of the disk, while accreted halo stars of the same age are consistently more metal poor.
This prediction can be directly tested once stellar ages are available for the \emph{Gaia} stars.}
\label{fig:ages}
\end{center}
\end{figure}

Ages of individual stars are another important diagnostic of their origin, so we now explore correlations between metallicity and age for stars in the Solar neighborhood, as predicted by the Latte simulation.
Figure~\ref{fig:ages} shows how the metallicity of stars in Latte depends on their age for disk (red), in-situ halo (light blue) and accreted halo (dark blue).
Shaded regions correspond to the 16th to 84th percentile in the distribution of metallicities for stars of a given age.
In general, metallicity increases with time, however, accretion of a significant amount of the low metallicity gas during the last significant merger 7\;Gyr ago is evident as a decrease in the metallicity of stars formed in situ immediately following this event.
Comparing the different populations in the simulation, we note that the metallicities of halo stars formed in situ closely follow the evolution of disk stars, while the accreted halo is more metal poor at all ages.
The bifurcation in the metallicity tracks for the in-situ and accreted halo is a prediction of the Latte simulation, which, if confirmed observationally, can be used to directly differentiate between the accreted and in-situ halo stars.

\section{Discussion}
\label{sec:discussion}
We have presented evidence that a substantial fraction the stellar halo in the solar neighborhood is metal-rich, and we now put this result in context of previous investigations (\S~\ref{sec:previous}).
Based on its chemical abundances and orbital properties, we inferred that the metal-rich halo has been formed in situ, so we discuss mechanisms that heat stars originally formed in the disk to halo-like orbits in \S~\ref{sec:diskheating}.
We conclude the section by outlining a test of the in-situ formation scenario for the metal-rich halo component with stellar ages in \S~\ref{sec:ages}.

\subsection{Previous evidence for a metal-rich halo}
\label{sec:previous}
Large-scale spectroscopic surveys have mapped the distribution of stellar metallicity in the Milky Way \citep[e.g.,][]{ivezic2008}.
Some of these stars have distance estimates, and can therefore be spatially identified with either the disk or the halo.
Relevant for this discussion, halo stars in the outer Galaxy, $R_{G}>15$\;kpc, are very metal-poor, with median $\rm[Fe/H]\sim-2.2$, while in the inner Galaxy, the typical metallicity of a halo star is $\rm[Fe/H]\sim-1.6$ \citep[e.g.,][]{carollo2007, dejong2010}, which agrees well with the metal-poor halo component identified in our sample.
Also similar to our findings, metallicities of the inner halo extend to the solar value, even though these are the tail of the metallicity distribution \citep[e.g.,][]{allendeprieto2006}, and not an additional component that is visible in Figure~\ref{fig:mdf}.

However, employing only spatial information makes it hard to distinguish between the halo and the thick-disk, especially at high metallicity.
The difference between the thick disk and the halo is more pronounced in kinematics, so \citet{sheffield2012} selected outliers from the disk velocity field to study the halo.
They identified eight metal-rich halo stars, whose $\alpha$-abundances are consistent with disk, and suggested these have been kicked out of the disk.
Still, none of these stars have kinematics that could rule out a thick disk interpretation at a $3\;\sigma$ level.

To date, only one metal-rich star has a high probability of being a halo star \citep{hawkins2015}.
This star has high velocity in the Galactic rest frame, $V_{GSR}\simeq430$\;km/s, and is on an eccentric orbit, $e=0.72$, that reaches $\sim30$\;kpc above the Galactic plane, but has a metallicity $\rm[Fe/H]=-0.18$, so \citet{hawkins2015} concluded it has been ejected from the thick disk.
This star is a valuable indicator of the processes governing the assembly of the Galaxy, but being a single star, it cannot establish how ubiquitous these processes are.

\subsection{Disk heating mechanisms}
\label{sec:diskheating}
The excess of metal-rich halo stars on prograde orbits indicates they originate from within the Milky Way disk, but it is unclear at what Galactocentric distance these stars were formed.
The bulk of similarly selected stars in Latte originated from the inner Galaxy, implying that a degree of radial migration and/or heating occurred.
Alternatively, these stars could be runaway stars -- stars kicked out of the disk at the Solar radius during dynamical processes not captured within the Latte simulation.
In this section we explore implications of these opposing disk heating mechanisms, and assess how plausible they are in explaining metal-rich halo stars detected in RAVE-on--TGAS.

\subsubsection{Runaway stars}
\label{sec:runaway}
Runaway stars are young stars that were formed in the disk and ejected from their birthplace \citep{blaauw1961}.
Some of them have been found in the halo \citep[e.g.,][]{conlon1990}, so it is sensible to test whether any of our metal-rich halo stars are in fact runaways.
Having formed in the disk recently, runaways should have high metallicities, and \citet{bromley2009} have already suggested that solar-metallicity stars reported at 5\;kpc away from the Galactic place \citep{ivezic2008} could be runaway stars.
The metallicity distribution of our metal-rich halo peaks at $\rm[Fe/H]\approx-0.5$, so most of them are probably not runaway stars.
However, the high-metallicity tail of our halo sample extends to super-solar values, so we test whether any of those stars are consistent with being runaways.

In addition to being metal-rich, runaway stars typically have an early spectral type, so to test the runaway origin of the metal-rich halo, we analyze the fraction of runaways expected in a disk population as a function of temperature.
Due to difficulties in the spectroscopic analysis of hot stars, the RAVE-on catalog preferentially contains cold stars, with more than 80\% of the sample being K stars.
The expected fraction of runaway stars drops sharply with decreasing temperature from 40\% for O stars, 5\% for B stars, $\approx2\%$ for A stars to negligible contributions from later stellar types \citep{blaauw1961, gies1986, bromley2009, perets2012}.
There are no OB stars in our halo sample, and only three A stars.
These all have a super-solar metallicity, $\rm[Fe/H]>0$, so they are prime candidates for runaway stars.
However, they constitute only 0.5\% of the metal-rich halo sample, so we can safely conclude that runaway stars are a minor component of the observed metal-rich stellar halo.

\subsubsection{Radial migration}
\label{sec:migration}
In the classical picture of radial migration, spiral structures in the disk radially scatter nearby stars up to several kpc \citep{sellwood2002}.
Stars that have migrated outwards are typically more metal-rich than their neighbors, so radial migration explains why some stars in the Solar neighborhood (possibly including the Sun!) are more metal-rich than both the surrounding stellar population and the local interstellar medium \citep{wielen1996}.
Idealized simulations of disk formation have found radial migration to be a key component in explaining numerous observables, such as the spread in the age--metallicity relation \citep{roskar2008} or the disk morphology and its abundance patterns \citep{schonrich2009}.
Such studies have also shown that stars migrating outwards reach larger heights above the disk plane \citep[e.g.,][]{schonrich2009, loebman2011}.
This scenario can explain, at least in part, the formation of the thick disk \citep[e.g.,][]{wilson2011}, so it is also conceivable that the metal-rich stars we identified on the halo-like orbits are an endpoint of this process.
However, subsequent numerical works have found that outward migrators do not attain larger scale heights from the disk plane \citep{minchev2012, vera-ciro2014}, thus casting doubts on the idea of forming the thick disk and a metal-rich halo through radial migration.

The above results on radial migration are based on simulations initialized to match the orderly morphology of stellar disks observed at the present day and do not capture the range of dynamical conditions present in the cosmological simulations of galaxy formation \citep[e.g.,][]{agertz2009}.
In particular, the metal-rich halo in the Latte simulation was formed at early cosmic times ($z \gtrsim 1$, or age $\gtrsim 7$ Gyr), before the formation of the thin disk \citep{ma2016} while the host galaxy was still actively accreting.
Accretion from the inter-galactic medium and merging satellites brought in a lot of gas to the galactic center (evident as star-forming tracks in Figure~\ref{fig:dform}), which fueled additional in-situ star formation.
\citet{elbadry2016} showed that these conditions create two mechanisms for radially displacing stars from their birth location.
First, some stars are formed during gas outflows, so their initial orbits can be eccentric and have large apocenters.
Second, the combination of inflowing gas accretion and gas outflows driven by stellar feedback produce strong fluctuations in the underlying gravitational potential.
Such fluctuations have already been shown to change the distribution of dark matter, in particular, in generating a cored density profile \citep[e.g.,][]{pontzen2012, brooks2014, dicintio2014, chan:fire.dwarf.cusps}, but \citet{elbadry2016} further showed that stellar orbits are affected as well, ultimately becoming heated to a more isotropic distribution.
This mechanism is most efficient in relatively shallow potential wells of dwarf galaxies.
While the Milky Way-mass host galaxy in Latte is too massive to exhibit such behavior at $z \sim 0$, its progenitor has significantly lower stellar mass (was more gas rich) while the metal-rich halo was forming, so we suggest that a similar process drove the radial migration to the Solar circle that we see here, consistent with conclusions of \citet{ma2017}.
We will investigate this behavior in Latte in more detail in future work.

Radial migration, driven by large-scale motions in the Milky Way progenitor, could explain the origin of metal-rich stars on halo-like orbits in the Solar neighborhood.
If these stars truly originate from the inner Galaxy, then they not only illustrate an important dynamical mechanism shaping the Galaxy, but are also a unique window into star formation in the early Milky Way.
Future analysis and data from the \emph{Gaia} mission will allow us to test this hypothesis, which we briefly discuss in the next section.

\subsection{Inferring the formation scenario of the stellar halo with stellar ages}
\label{sec:ages}
In section \S\ref{sec:formation}, we showed that the metal-rich halo simulated in Latte was chemically enriched in a fashion similar to the Latte disk.
When compared at a fixed age, these components have higher metallicity than the metal-poor halo at all times (see Figure~\ref{fig:ages}).
We can directly test this prediction by dating the stars in our TGAS--RAVE-on sample.
Unfortunately, stellar ages are not obtained easily \citep[for a recent review, see][]{soderblom2010}.
A number of observables that correlate with age have been identified, such as stellar rotation \citep{barnes2007}, chromospheric activity \citep{mamajek2008}, or surface abundances \citep{ness2016}, but none of these empirical relations are applicable to all of the field stars.
Models of stellar evolution can relate the position of any star in the Hertzsprung--Russell diagram (HRD) and its internal structure to its age.
The latter is inferred from asteroseismic studies of stellar pulsations, and has so far been employed to date a few dozen of well observed stars \citep[e.g.,][]{keplerages}.
In the coming decade, asteroseismic dating will be expanded, but still limited to the brightest stars \citep{tess, plato}.
We expect the HRD age dating to be more easily applied to a larger sample of stars, and discuss it in more detail below.

Coeval stellar populations are routinely dated by comparison of their tracks in the HRD to theoretical isochrones \citep[e.g.,][]{sandage1970, chaboyer1998, dotter2007}, but isochrone dating of field stars is less straightforward.
Intrinsically, without the HRD positions of coeval companions, age estimates of field stars are very uncertain in evolutionary stages which keep stars at an approximately constant position in the HRD, such as the main sequence phase.
In addition, precisely measuring stellar distances, which are required to put a star on the HRD, as opposed to merely on a color-magnitude diagram, is observationally challenging.
However, if distances are known, stellar ages can be measured for stars in pre- or post-main sequence evolutionary stages.
\citet{gcs} measured ages and other intrinsic stellar parameters for thousands of nearby field stars by obtaining their absolute magnitudes from Hipparcos parallaxes, effective temperatures and metallicities from follow-up spectroscopy, and then reading off the age by interpolating theoretical isochrones in this three-dimensional space.
TGAS has already increased the sample of stars with known distances by an order of magnitude, and several groups are modeling the multi-band stellar photometry (and including spectroscopy when available) to provide constraints on their ages.
In such a procedure, ages of red giants are measured with a precision of $1-3$\;Gyr when only photometric data is available.
Including spectroscopically derived stellar parameters reduces the uncertainty in recovered ages to 1\;Gyr (P.~Cargile, private communication).
Most of the halo stars in our sample are giants, so if the bifurcation in the age--metallicity relation of halo stars exists at the level suggested by the simulation studied here, we will soon be able to detect it observationally.

\section{Summary}
\label{sec:summary}
The goal of this work was to investigate the origin of halo stars in the Solar neighborhood.
We analyzed a sample of stars located in 6D using the first year of \emph{Gaia} data, combined with observations from ground-based spectroscopic surveys.
The halo was defined by requiring a relative velocity with respect to the Local Standard of Rest larger than 220\;km/s.
Metal-poor stars, $\rm[Fe/H]<-1$, comprise approximately half of this kinematically defined halo sample.
The other half of the sample are stars more metal-rich than expected for the inner halo ($\rm[Fe/H]>-1$), and whose metallicity and $\alpha$-element abundances are traditionally associated with the disk population.
We built a toy model of the Solar neighborhood to show that orbits of the metal-poor halo stars are isotropically distributed, while the metal-rich halo is preferentially aligned with the Galactic disk.
To uncover the origin of these two halo components, we performed an identical analysis on a sample of stars selected in a Solar-like neighborhood of the Latte cosmological hydrodynamic simulation.
A significant fraction of the simulated halo stars are also metal-rich and on prograde orbits with respect to the disk, and most of them were formed in situ in the main galaxy.
In general, we found that the origin of halo stars is well correlated with stellar metallicity, with metal-poor stars having been accreted, and metal-rich having formed in situ, and has little dependence on their kinematics.
Additionally, the majority of metal-rich halo stars in Latte were formed in the inner galaxy, and migrated several kpc outwards, indicating that radial redistribution actively operates in Milky Way-like galaxies.
In this model, the in-situ component of the stellar halo is more metal-rich than the accreted component at a fixed age, a hypothesis easily verifiable with forthcoming data.

We have presented for the first time a large population of metal-rich, $\rm[Fe/H]>-1$, stars on halo-like orbits, and inferred that they have formed in situ.
These conclusions arise in part from a remarkable agreement between the precise observational data from the \emph{Gaia} mission and the Latte simulation of a Milky Way-like galaxy.
Direct comparison of observed and simulated galaxies at this level of detail will be a powerful tool in studying the Galaxy in the \emph{Gaia} era.

\begin{figure}
\begin{center}
\includegraphics[width=\columnwidth]{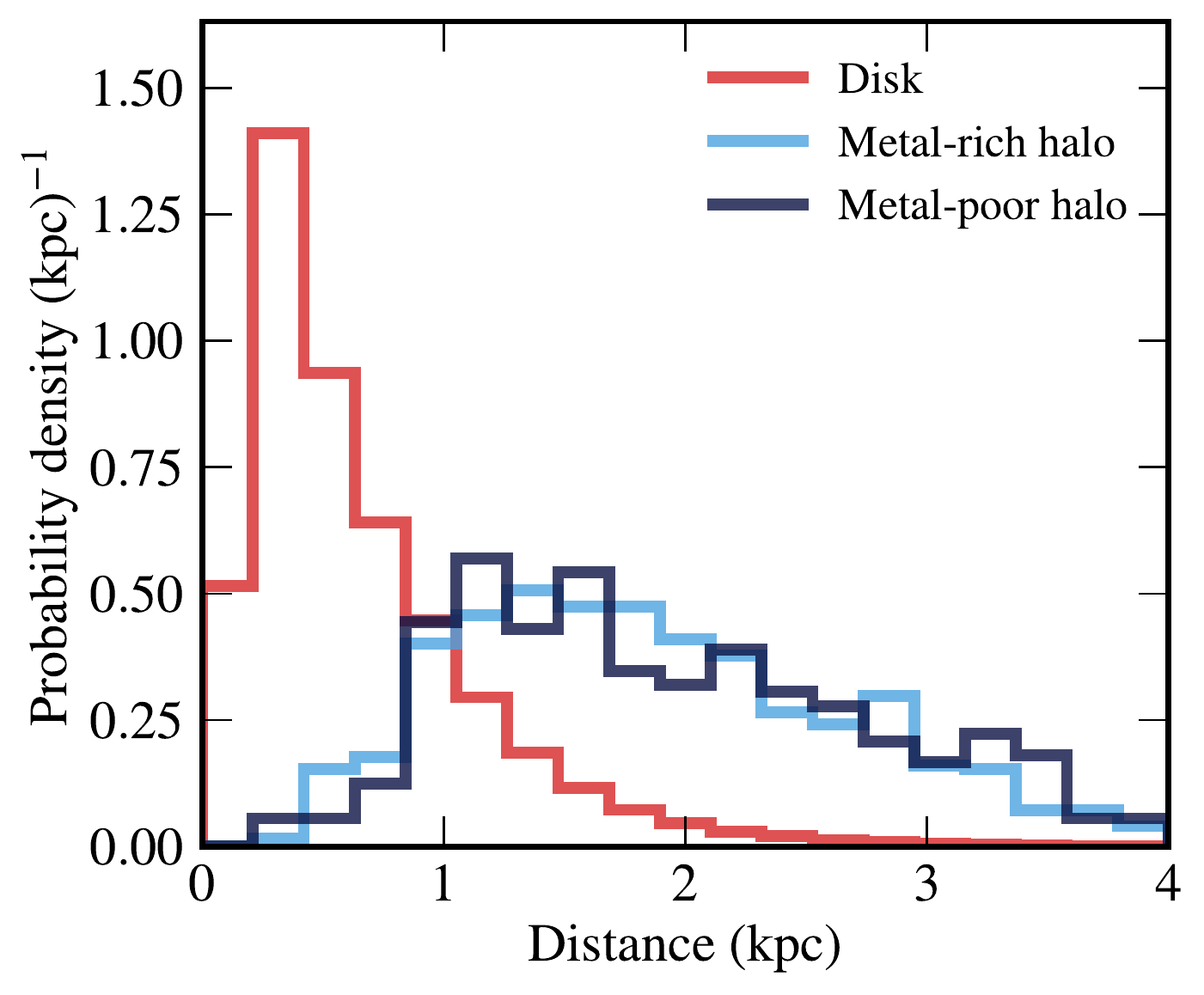}
\caption{Distribution of heliocentric distances in the TGAS--RAVE-on sample, split by kinematically identified components.
The disk stars (red) are in general closer than the halo (blue), but we see no difference in the distances of the metal-rich (light blue) and metal-poor halo component (dark blue).
Orbital differences observed between the metal-rich and the metal-poor halo do not originate from a difference in their spatial distributions.
}
\label{fig:distance}
\end{center}
\end{figure}

\vspace{0.5cm}
\emph{Acknowledgments:}
It is a pleasure to thank Andy Casey for providing a match of the RAVE-on catalog to TGAS, Yuan-Sen Ting for matching the APOGEE catalog to TGAS, Kim Venn, Rosy Wyse, Warren Brown and Elena D'Onghia for insightful comments that shaped the progression of this project.

This work has made use of the following Python packages: \texttt{matplotlib} \citep{mpl}, \texttt{numpy} \citep{numpy}, \texttt{scipy} \citep{scipy}, \texttt{Astropy} \citep{astropy} and \texttt{gala} \citep{gala}.

This paper was written in part at the 2016 NYC Gaia Sprint, hosted by the Center for Computational Astrophysics at the Simons Foundation in New York City.

AB was supported by an Institute for Theory and Computation Fellowship.
CC acknowledges support from the Packard Foundation.
AW was supported by a Caltech-Carnegie Fellowship, in part through the Moore Center for Theoretical Cosmology and Physics at Caltech, and by NASA through grant HST-GO-14734 from STScI.
DK was supported by NSF grant AST-1412153 and a Cottrell Scholar Award from the Research Corporation for Science Advancement.

This work has made use of data from the European Space Agency (ESA) mission {\it Gaia} (\url{http://www.cosmos.esa.int/gaia}), processed by the {\it Gaia} Data Processing and Analysis Consortium (DPAC, \url{http://www.cosmos.esa.int/web/gaia/dpac/consortium}). Funding for the DPAC has been provided by national institutions, in particular the institutions participating in the {\it Gaia} Multilateral Agreement.

Funding for RAVE has been provided by: the Australian Astronomical Observatory; the Leibniz-Institut fuer Astrophysik Potsdam (AIP); the Australian National University; the Australian Research Council; the French National Research Agency; the German Research Foundation (SPP 1177 and SFB 881); the European Research Council (ERC-StG 240271 Galactica); the Istituto Nazionale di Astrofisica at Padova; The Johns Hopkins University; the National Science Foundation of the USA (AST-0908326); the W. M. Keck foundation; the Macquarie University; the Netherlands Research School for Astronomy; the Natural Sciences and Engineering Research Council of Canada; the Slovenian Research Agency; the Swiss National Science Foundation; the Science \& Technology Facilities Council of the UK; Opticon; Strasbourg Observatory; and the Universities of Groningen, Heidelberg and Sydney.
The RAVE web site is at \url{https://www.rave-survey.org}.

Funding for the Sloan Digital Sky Survey IV has been provided by the Alfred P. Sloan Foundation, the U.S. Department of Energy Office of Science, and the Participating Institutions. SDSS-IV acknowledges support and resources from the Center for High-Performance Computing at the University of Utah. The SDSS web site is \url{www.sdss.org}.

SDSS-IV is managed by the Astrophysical Research Consortium for the Participating Institutions of the SDSS Collaboration including the Brazilian Participation Group, the Carnegie Institution for Science, Carnegie Mellon University, the Chilean Participation Group, the French Participation Group, Harvard-Smithsonian Center for Astrophysics, Instituto de Astrof\'isica de Canarias, The Johns Hopkins University, Kavli Institute for the Physics and Mathematics of the Universe (IPMU) / University of Tokyo, Lawrence Berkeley National Laboratory, Leibniz Institut f\"ur Astrophysik Potsdam (AIP),  Max-Planck-Institut f\"ur Astronomie (MPIA Heidelberg), Max-Planck-Institut f\"ur Astrophysik (MPA Garching), Max-Planck-Institut f\"ur Extraterrestrische Physik (MPE), National Astronomical Observatories of China, New Mexico State University, New York University, University of Notre Dame, Observat\'ario Nacional / MCTI, The Ohio State University, Pennsylvania State University, Shanghai Astronomical Observatory, United Kingdom Participation Group, Universidad Nacional Aut\'onoma de M\'exico, University of Arizona, University of Colorado Boulder, University of Oxford, University of Portsmouth, University of Utah, University of Virginia, University of Washington, University of Wisconsin, Vanderbilt University, and Yale University.

\begin{figure*}
\begin{center}
\includegraphics[width=\textwidth]{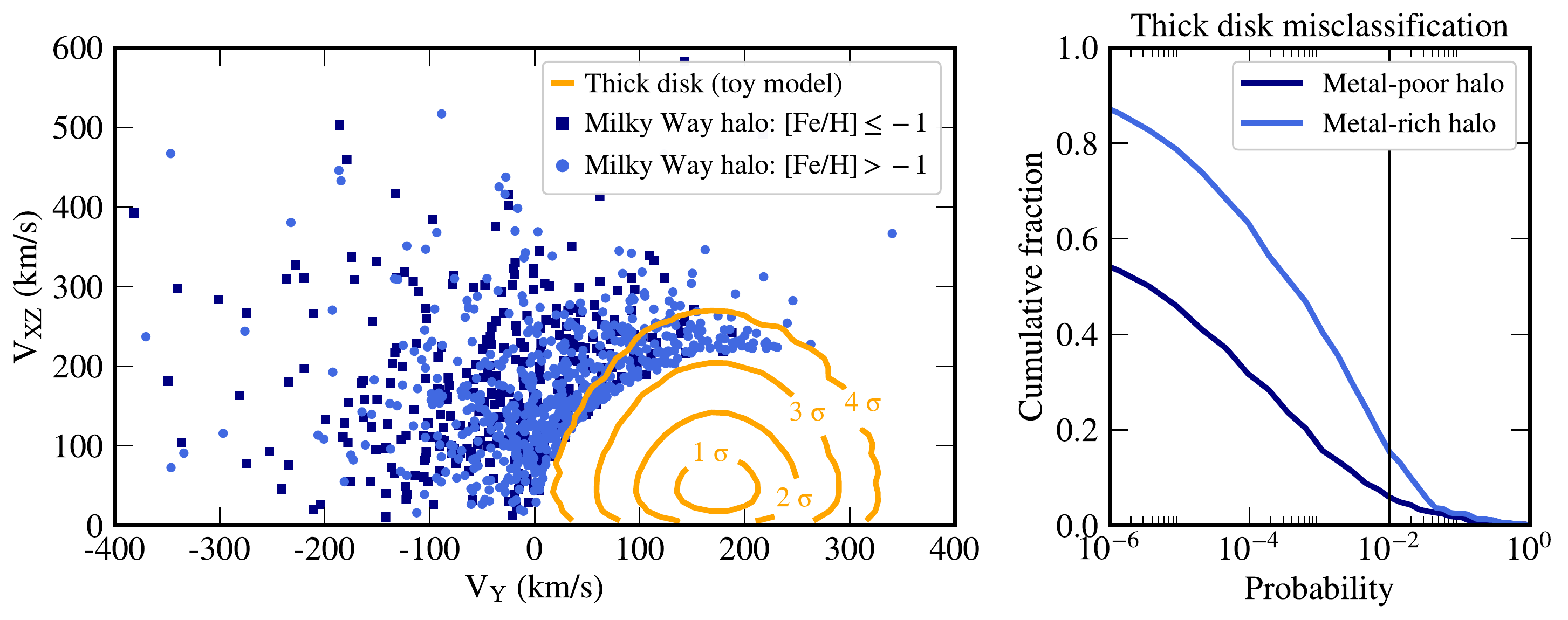}
\caption{(Left) Probability contours of toy model thick disk stars in the Toomre diagram in whole steps of standard deviation, $\sigma$ (orange lines).
All halo stars from our RAVEon--TGAS sample (metal-rich in light blue circles and metal-poor in dark blue squares) lie outside of the $3\;\sigma$ thick disk contour, but some are consistent with the thick disk at a $4\;\sigma$ level.
(Right) Probability for stars, identified in RAVEon--TGAS as part of the halo, of actually being a part of the thick disk.
Lines show cumulative fractions of halo stars as a function of this probability, with light blue for the metal-rich and dark blue for the metal-poor halo stars.
Only a small fraction of both halo components is expected to be a misclassified part of the thick disk (20\% of the metal-rich and 5\% of the metal-poor halo have a thick disk probability larger than 1\%, marked with a black vertical line).}
\label{fig:tdcont}
\end{center}
\end{figure*}

\bibliographystyle{apj}
\bibliography{apj-jour,mrich_halo}

\appendix{}
\section{Distribution of distances in the TGAS--RAVE-on sample}
\label{sec:distances}
The TGAS--RAVE-on sample employed in this work is exploring the Solar neighborhood, but is not volume complete.
To illustrate the extent of the sample, we show the distribution of heliocentric distances in Figure~\ref{fig:distance}, and compare the distances of the kinematically selected components defined in \S\ref{sec:sample}.
This sample is local in the sense that more than 90\% of the stars are within 1.5\;kpc from the Sun.
As expected from our position close to the Galactic plane, the disk stars (red line) are preferentially closer than the halo stars (blue lines).
90\% of the disk is contained within 1.3\;kpc, while the 90th percentile in the halo distance extends to 3\;kpc.
On the other hand, the two halo components split by metallicity have a similar distribution of distances and are spatially indistinguishable within this volume.

\section{Thick disk contamination}
\label{sec:tdcontamination}
The thick disk bridges the thin disk and the halo in both chemical abundances and kinematics.
Given how the metal-rich halo identified in this study has abundances consistent with the thick disk, in this section we quantify how different it is from the canonical thick disk kinematically.

As demonstrated by the toy model of the Solar neighborhood (\S\ref{sec:toymodel}), we expect some thick disk stars to enter our halo selection (Figure~\ref{fig:toy}, red points above the thick black line in the left panel).
We visualize the expected contamination levels in the left panel of Figure~\ref{fig:tdcont} by drawing the probability contours for the thick disk velocity ellipsoid \citep{bensby2003} in the Toomre diagram.
The successive contours enclose the parameter space occupied by the thick disk with probabilities of 68\%, 95\%, 99.7\% and 99.9\% (labeled as $1-4\;\sigma$ in Figure~\ref{fig:tdcont}).
Halo stars from our sample are shown as points, with metal-rich being represented by light blue circles and metal-poor by dark blue squares.
All of the halo stars are outside of the $3\;\sigma$ thick disk contour, or inconsistent with being a thick disk at the 99.7\% level, but 145 ($\approx25\%$) metal-rich and 25 ($\approx7\%$) metal-poor halo stars are inside the $4\;\sigma$ contour.
On the other hand, there are 453 metal-rich halo stars outside the $4\;\sigma$ thick disk contour, while only 16 thick disk stars are are expected in this region by the toy model.
Even though the velocity distributions of the halo and the thick disk are overlapping, and our halo sample may not be completely pure of the thick disk contaminants, there is a clear excess of stars with thick disk abundances beyond the canonical thick disk kinematics.

In the right panel of Figure~\ref{fig:tdcont} we quantify the probability of halo stars in our sample being a part of the thick disk, fully accounting for the observational uncertainties in all six observables.
The light blue line shows the cumulative fraction of metal-rich halo stars being a thick disk star at a given probability, while dark blue is the corresponding line for the metal-poor halo stars.
Less than 20\% of metal-rich halo stars have more than a percent probability (marked by a vertical black line) of being a misclassified thick disk star.
For the metal-poor halo, this fraction is even lower at 5\%.
The median probability of being a thick disk star is $\sim3\times10^{-4}$ and $\sim2\times10^{-6}$ for the metal-rich and the metal-poor halo, respectively, ruling out the thick disk interpretation of metal-rich stars identified in the local stellar halo.

So far, we have only considered the kinematic definition of a thick disk as measured by \citet{bensby2003}.
Studies based on different samples have arrived at slightly modified properties of a thick disk velocity ellipsoid \citep[e.g.,][]{soubiran2003, carollo2010}.
Furthermore, in a theoretical study of a thick disk formed in an idealized simulation, \citet{sb2009} noted that its velocity distribution in the Toomre diagram is much more asymmetric than the usually assumed Gaussian distribution function.
However, even this more extended definition of a thick disk does not encompass all of the metal-rich stars identified in the Solar neighborhood, excluding in particular stars on retrograde orbits with high $V_{XZ}$.
Assuming a different distribution function for the thick disk changes the inferred contamination levels in our halo sample in detail, but no disk-like distribution explains stars on very retrograde, kinematically hot orbits, where some of the metal-rich stars from our sample are found.

\end{document}